\documentclass[11pt,a4paper,nofootinbib,superscriptaddress]{revtex4-1}

\pdfoutput=1

\usepackage[utf8]{inputenc}
\usepackage{graphicx,bm,bbm}
\usepackage{amssymb,graphicx,epstopdf}
\usepackage{slashed,subfigure}
\usepackage{caption}
\usepackage{xcolor}
\usepackage{amsmath,mathtools}
\usepackage{epstopdf,dcolumn}
\usepackage{soul}
\usepackage{mathtools}
\allowdisplaybreaks
\usepackage[normalem]{ulem}
\usepackage{cancel}
\usepackage{xcolor}
\usepackage[colorlinks=true,linktocpage=true,linkcolor=blue,citecolor=blue]{hyperref}
\usepackage{xcolor}
\usepackage{braket}
\allowdisplaybreaks
\usepackage{enumitem}
\usepackage{mathrsfs}

\def\be {\begin{equation}}
\def\ee {\end{equation}}
\def\bea {\begin{eqnarray}}
\def\eea {\end{eqnarray}}
\def\bc {\begin{center}}
	\def\ec {\end{center}}
\def\nn {\nonumber}

\def\om{\omega}
\def\({\left(}
\def\){\right)}
\def\[{\left[}
\def\]{\right]}

\DeclareGraphicsExtensions{.jpg,.pdf,.eps}

\begin{document}

	
\title{Magnetic field-dependent electric charge transport in hadronic medium at finite temperature}	

	\author{Ritesh Ghosh}
	\email{riteshghosh1994@gmail.com}
	\affiliation{School of Physical Sciences, National Institute of Science Education and Research, An OCC of Homi Bhabha National Institute,\\  Jatni, Khurda 752050, India}
		\affiliation{
		Theory Division, Saha Institute of Nuclear Physics, A CI of Homi Bhabha National Institute, 1/AF, Bidhannagar, Kolkata 700064, India}

	\author{Manu Kurian}
	\email{manu.kurian@mail.mcgill.ca}
	\affiliation{Department of Physics, McGill University, 3600 University Street, Montreal, QC, H3A 2T8, Canada}

	\begin{abstract}
Electric charge transport of hadronic matter at finite temperature and magnetic field is studied within the linear sigma model. Anisotropic transport coefficients associated with the charge transport are estimated both in the weak and strong regimes of the magnetic field using the transport theory approach. In a weakly magnetized medium, the magnetic field effects are incorporated through the Lorentz force term in the Boltzmann equation. Strong magnetic field puts further constraints on the motion of charged particles through Landau quantization. Magnetic field-dependent thermal relaxation time is obtained from interaction rates of hadrons with the S-matrix approach by considering the Landau level kinematics of the charged hadrons. Mean-field effects are embedded in the analysis through the temperature-dependent hadron masses. Further, the hadronic medium response to a time-varying external electric field is studied in weak and strong magnetic field regimes. It is seen that electromagnetic responses of the hadronic matter have a strong dependence on the mean-field effects, sigma mass, the strength of the external fields, and its evolution in the medium. 
	\end{abstract}
	
	\maketitle 
	\newpage
	
	\section{Introduction}
Research programs at Relativistic Heavy Ion Collider (RHIC) at Brookhaven National Laboratory (BNL) and Large Hadron Collider (LHC) at CERN have realized the existence of strongly interacting matter-the Quark-Gluon Plasma (QGP) at extreme conditions~\cite{Adams:2005dq, Back:2004je,Arsene:2004fa,Adcox:2004mh,Aamodt:2010pb}.  The dynamical evolution of the created medium has been successfully studied with hydrodynamic modelling~\cite{Gale:2013da, Romatschke:2009im, Heinz:2013th}. Transport coefficients of the strongly interacting matter act as the input parameters for the hydrodynamic simulation. The extraction of transport coefficients from effective microscopic theories and its phenomenological constraints from the measured observables in the collision experiments at the RHIC  and LHC are interesting aspects of current research in heavy-ion collision physics~\cite{Das:2022lqh}. 

Several theoretical efforts~\cite{Kharzeev:2007jp, Skokov:2009qp, Voronyuk:2011jd,Deng:2012pc,Zhong:2014cda} have suggested the existence of a strong magnetic field in a non-central heavy-ion collision in which the magnitude of the field can be in the order of $10^{18} $ G at the RHIC and  $10^{19} $ G at the LHC. The recent observations on the directed flow of neutral D mesons at the RHIC and LHC~\cite{Acharya:2019ijj,Adam:2019wnk} gave indications of the generation of a strong magnetic field in the initial stages of the collision experiment. However, the space-time evolution of the magnetic field is not completely understood so far, and some recent analysis have shown that the field may persist longer time in the conducting medium than expected~\cite{Tuchin:2013bda,Inghirami:2016iru}. The significance of the generated magnetic field has been studied in the context of novel phenomena such as chiral magnetic effect~\cite{Fukushima:2008xe}, chiral vortical effect~\cite{Kharzeev:2015znc}, and in the realization of global hyperon polarization~\cite{Becattini:2016gvu}. Properties of quarkonia~\cite{ Ghosh:2022sxi,Singh:2017nfa}, magnetic catalysis~\cite{Mueller:2015fka}, inverse magnetic catalysis~\cite{Endrodi:2019zrl, DElia:2018xwo}, QCD thermodynamic and transport properties~\cite{Kurian:2018qwb,Bandyopadhyay:2017cle,Karmakar:2019tdp, Kurian:2017yxj, Panday:2022rgv, Kurian:2020qjr,Khan:2022orx,K:2022pzc}, chiral susceptibility~\cite{Ghosh:2021knc}, dilepton production~\cite{Das:2021fma, Bandyopadhyay:2016fyd}, damping rate of photon~\cite{Ghosh:2019kmf} have been investigated in presence of a magnetic field. The impact of electromagnetic fields on the conducting medium can be studied in terms of electric charge transport and the associated conductivities that quantify the induced current due to the external fields. Several efforts have been done to explore the behaviour of electrical conductivity of the QCD medium within the Kubo formalism~\cite{Feng:2017tsh, Satapathy:2021cjp, Hattori:2016lqx}, lattice QCD approach~\cite{Buividovich:2010tn}, holographic models~\cite{Pu:2014fva}, and transport theory calculations~\cite{Satow:2014lia, Gorbar:2016qfh, Gowthama:2020ghl, K:2021sct, Kerbikov:2014ofa, Ghosh:2019ubc,Das:2019wjg,Kurian:2019fty}.  The electromagnetic responses of the QCD medium have a crucial role in magnetohydrodynamics simulations and in the study of the impact of the fields on the measured observables in the heavy-ion collision experiments.

The present study is on the electric charge transport in a magnetized hadronic medium by employing the linear sigma model (LSM). The LSM, first introduced by Gell-Mann and Levy~\cite{Gell-Mann:1960mvl} is a simple, low-energy effective model to study the hadronic system. Later, LSM have been extended by including quarks~\cite{Ayala:2018zat,Das:2019ehv} and vector mesons in this model~\cite{Divotgey:2016pst}. In the presence of an external magnetic field,  pion condensate~\cite{Loewe:2013coa}, chiral phase transition~\cite{Ayala:2009ji}, and neutral pion mass~\cite{Das:2019ehv} have been analyzed using the LSM. The LSM description of transport coefficients of the hadronic medium has been presented within the relaxation time approximation   ~\cite{Chakraborty:2010fr}. Recently, in Ref.~\cite{Heffernan:2020zcf}, the behaviour of shear viscosity, bulk viscosity, and electrical conductivity of the hadronic matter has been studied using a functional variational approach. The prime focus of the current analysis is on the effective description of the hadronic medium response to electromagnetic fields. As the strength of the generated magnetic field is not completely known throughout the medium evolution, the analysis has been done both in the weak and strong magnetic field regimes. In a weakly magnetized matter, the temperature is the dominant energy scale, whereas the magnetic field act as a small perturbation in the medium. On the other hand, in the strongly magnetized hadronic matter, the charged particles follow Landau level kinematics. In both cases, we employ the transport theory approach within the relaxation time approximation and obtain the magnetic field-dependent conductivities in a hadronic medium with the LSM for the first time. Unlike in the case of a weakly magnetized medium, the impact of the strong magnetic field on thermodynamics and interaction frequency has been considered in the analysis. Several recent studies have suggested the existence of space-time decaying electromagnetic fields in the collision experiments~\cite{Deng:2012pc,Tuchin:2013apa,Hongo:2013cqa}. This sets the motivation to extend the analysis to explore the hadronic medium response to an external time-varying electric field. 
 The impact of the proper time evolution of the external electric field on the charge transport is studied in a weakly and strongly magnetized hadronic matter. We have illustrated the effects of mass of $\sigma$ meson, strength, and evolution of external fields on the induced current density in a magnetized hadronic medium.

The article is organized as follows. The general formalism for the electric charge transport process in a weakly magnetized hadronic medium followed by the LSM description of the medium is presented in Sec.~\ref{EW}. In Sec.~\ref{ES}, the longitudinal current density of the hadronic matter in the presence of a strong magnetic field and the magnetic field-dependent interaction frequency is discussed. Sec.~\ref{NE} is devoted to the description of medium responses to a time-varying electric field in both weakly and strongly magnetized hadronic matter. We presented the results in Sec.~\ref{RE} and finally summarised the analysis in Sec.~\ref{SU}. 

{{\it{ Notations and conventions}:}    
The particle velocity is denoted as ${\bf v}=\frac{{\bf p}}{\epsilon}$ with ${\bf p}$ and $\epsilon$ as the momentum and energy, respectively. 
The components of a vector ${\bf A}$ are represented with $A^l$. The quantities $B = |{\bf B}|$, $E = |{\bf E}|$, and $j = |{\bf j}|$ describes the magnitude of the magnetic field, electric field and current density in the medium, respectively. The metric tensor is taken as $g^{\mu\nu}=\text{diag}(1, -1, -1, -1)$.}

\section{Electric charge transport of hadronic matter in a weak uniform magnetic field}\label{EW}
Electric field acts as the source of perturbation associated with the charge transport of the hadronic medium and can be quantified in terms of induced current density, ${\bf j}=\sigma_0 {\bf E}$ with $\sigma_0$ as the electrical conductivity. In the presence of a magnetic field, the motion of the charged particle will be constrained and this induces anisotropy in the transport processes in the medium. The anisotropic momentum transport coefficients (five shear-viscous coefficients and two components of bulk viscosity) in the QCD medium have been studied in  Refs.~\cite{Dash:2020vxk,Ghosh:2020wqx,Panda:2020zhr,Huang:2011dc,Denicol:2019iyh,Rath:2022oum}. Depending upon the strength of the magnetic field, the electric charge transport and the decomposition of the current density will be modified. In the presence of a weak magnetic field, the electric current density can be defined as,
\begin{align}\label{I.1}
j^l=\sigma_{0} \delta^{lj} E_j -\sigma_1 \epsilon^{lkj}b_kE_j+\sigma_2 b^lb^jE_j.
 \end{align}
where $\sigma_0$, $\sigma_1$, and $\sigma_2$ are the transport coefficients associated with the electric charge transport in the presence of constant electromagnetic fields. Here, $\delta^{lj}$ denotes the Kronecker delta function, $\epsilon^{lkj}$ is the antisymmetric $3 \times 3$ tensor, and ${\bf b}$ is the direction of the magnetic field in the hadronic matter. The current density can be defined in terms of particle momentum distribution function as follows,
\begin{align}\label{I.2}
j^l=\sum_a \int \frac{d^3 \textbf{p}}{(2\pi)^3}v^lq_a \delta f_a.
 \end{align}
where the subscript $a$ describes the particle species, $q_a$ is the charge and $\delta f_a$ is the non-equilibrium part of the distribution function that describes the near-equilibrium distribution of hadronic particles in the presence of external electromagnetic fields. The first step in the calculation of current density is to estimate $\delta f_a$ by solving the relativistic transport equation. In the presence of a weak magnetic field, the Boltzmann equation within the relaxation time approximation takes the form as follows,
\begin{align}\label{I.3}
{p}^{\mu}\,\partial_{\mu}f_a(x,{p})+q_{a}F^{\mu\nu}{p}_{  \nu}\partial^{(p)}_{\mu} f_a=-\frac{\delta f_a}{\tau_a},
\end{align}
in which $\tau_a$ is the thermal relaxation time, $F^{\mu\nu}$ is the field strength tensor and $f_a=f^0_a+\delta f_a$ is the near-equilibrium distribution function of the hadronic particles where $f^0_a$ is the equilibrium distribution function. The following ansatz is considered to estimate the non-equilibrium part of the momentum distribution, 
\begin{align}\label{I.4}
\delta f_a= {\bf{p}}\cdot\Big[\alpha_1\textbf{E}+\alpha_2 \textbf{B} + \alpha_3(\textbf{E}\times \textbf{B})\Big] \frac{\partial f^0_a}{\partial \epsilon}, 
\end{align}
where $\alpha_1$, $\alpha_2$, $\alpha_3$ are the unknown functions that can be obtained from the microscopic description of the hadronic matter. By employing the form of $\delta f_a$ in the Boltzmann equation, we have
\begin{align}\label{I.5}
q_a{\bf v}_a\cdot\Big(
{\bf E}+ \alpha_1  ({\bf B}\times {\bf E})+\alpha_3  ({\bf E}\cdot{\bf B}){\bf B}-\alpha_3B^2{\bf E}\Big)=&-\frac{\epsilon_a}{\tau_{a}}\Big[\alpha_1{\bf v}_a\cdot{\bf E}+\alpha_2{\bf v}_a\cdot{\bf B}+\alpha_3{\bf v}_a\cdot({\bf E}\times{\bf B})\Big].
\end{align}
By comparing the independent tensorial structures such as $({\bf v}_a\cdot{\bf E})$, $({\bf v}_a\cdot{\bf B})$, and $({\bf v}_a\cdot({\bf E}\times{\bf B}))$ in both sides of Eq.~(\ref{I.5}), we obtain a set of equations that relate coefficients $\alpha_1$, $\alpha_2$, and $\alpha_3$ as,
\begin{align}\label{I.6}
& -\frac{\epsilon_a}{\tau_{a}}\alpha_1=q_a+\alpha_3q_{a} B^2,
&& \frac{\epsilon_a}{\tau_{a}}\alpha_2=\alpha_3q_{a}({\bf E}\cdot{\bf B}),
&&& \frac{\epsilon_a}{\tau_{a}}\alpha_3=\alpha_qq_{a}.
\end{align}
Employing the relation between the coefficients as described in Eq.~(\ref{I.6}), the form of $\alpha_1$, $\alpha_2$, and $\alpha_3$ can be obtained as follows,
\begin{align}\label{I.7}
&\alpha_1=-\frac{\epsilon_a q_a\tau_a}{(\tau_{a}^2 q_a^2B^2+\epsilon_a^2)},
&&\alpha_2=-\frac{( q_a\tau_a)^3({\bf E}\cdot{\bf B})}{(\tau_{a}^2 q_a^2B^2+\epsilon_a^2)\epsilon_a},
&&&\alpha_3=-\frac{ q^2_a\tau^2_a}{(\tau_{a}^2 q_a^2B^2+\epsilon_a^2)}.
\end{align}
Substituting Eq.~(\ref{I.4}) in Eq.~(\ref{I.2}), we obtain the electric charge current density in a weakly magnetized hadronic matter as,
\begin{align}\label{I.8}
{\bf j}=\frac{1}{3}\sum_a \int \frac{d^3 \textbf{p}}{(2\pi)^3} \frac{p^2}{\epsilon_a}q_a\bigg\{\frac{\epsilon_a q_a\tau_a}{(\tau_{a}^2 q_a^2B^2+\epsilon_a^2)}{\bf E}+\frac{( q_a\tau_a)^3({\bf E}\cdot{\bf B})}{(\tau_{a}^2 q_a^2B^2+\epsilon_a^2)\epsilon_a}{\bf B}+\frac{ q^2_a\tau^2_a}{(\tau_{a}^2 q_a^2B^2+\epsilon_a^2)}({\bf E}\times{\bf B})\bigg\}\Big(-\frac{\partial f_a^0}{\partial\epsilon_a}\Big).
\end{align}
The first term, which is proportional to the external electric field, represents the Ohmic current and is the leading order in relaxation time. The magnetic field in the medium induces other components of the current density depending upon the direction of the fields. For the case with transverse external fields, $i.e.$, ${\bf E}\cdot{\bf B}=0$, the second term in the Eq.~(\ref{I.8}) vanishes. { The Hall current density which is proportional to ${\bf E}\times{\bf B}$ will be subdominant in comparison with the Ohmic current as Hall component depends upon the chemical potential $\mu$.} Comparing Eq.~(\ref{I.1}) and Eq.~(\ref{I.8}), we define the transport coefficients associated with the charge transport in a weakly magnetized hadronic medium as,
\begin{align}
{\sigma_0}=\frac{1}{3}\sum_a \int \frac{d^3 \textbf{p}}{(2\pi)^3} {p^2}q^2_a\frac{\tau_a}{(\tau_{a}^2 q_a^2B^2+\epsilon_a^2)}\Big(-\frac{\partial f_a^0}{\partial\epsilon_a}\Big),\label{I.9}\\
 {\sigma_1}=\frac{1}{3}\sum_a \int \frac{d^3 \textbf{p}}{(2\pi)^3} \frac{p^2}{\epsilon_a}q^3_a\frac{\tau^2_aB}{(\tau_{a}^2 q_a^2B^2+\epsilon_a^2)}\Big(-\frac{\partial f_a^0}{\partial\epsilon_a}\Big),\label{I.10}\\
 {\sigma_2}=\frac{1}{3}\sum_a \int \frac{d^3 \textbf{p}}{(2\pi)^3} \frac{p^2}{\epsilon_a^2}q^4_a\frac{\tau^3_aB^2}{(\tau_{a}^2 q_a^2B^2+\epsilon_a^2)}\Big(-\frac{\partial f_a^0}{\partial\epsilon_a}\Big).\label{I.11}
\end{align}
The quantitative estimation of the conductivity coefficients required the knowledge of microscopic interactions of the hadronic medium. To that, one needs to obtain the thermal relaxation time associated with the interactions. In the current analysis, we ignored the effect of a weak magnetic field on the thermodynamic quantities and on the microscopic interactions in the hadronic medium as the temperature is the dominant scale in comparison with the strength of the field $qB\ll T^2$. However, magnetic has a strong dependence on the thermal relaxation time in the presence of a strong magnetic field $qB\gg T^2$ and is presented in section~\ref{LSMB}. Proper modeling of the hadronic medium is the first step towards the estimation of interaction frequency or relaxation time. 

\subsection*{{Linear sigma model} }
\label{sec:LSM}
	 
    The LSM is used in this work to calculate the electric conductivity. Mainly, the LSM Lagrangian contains bosonic field with $N$ components. When $N=4$, it denotes the theory of soft pion dynamics with $(N-1)$ pion fields $(\pi_i)$ and one sigma $(\sigma)$ field. The classic LSM Lagrangian density for $N=4$ takes the form as~\cite{Heffernan:2020zcf,Scavenius:2000qd},
	\bea
	\mathcal{L}&=&\frac{1}{2} (\partial_\mu \sigma)^2+\frac{1}{2} (\partial_\mu \boldsymbol{\pi} )^2-V(\sigma,\boldsymbol{\pi}).
	\eea
	The potential term in the above equation reads as,
	\bea
	V(\sigma, \boldsymbol{\pi})=\frac{\lambda}{4}(\sigma^2+\boldsymbol{\pi}^2-f^2)^2-H\sigma,
	\eea
	with $H \sigma$ represents the explicit chiral symmetry breaking term that describes the pion mass. The vacuum expectation value $v$ of the scalar $\sigma$ field  with $\sigma=v+\Delta$, where $\Delta$ is the fluctuation, is described as follows,
	\bea
	\lambda v (v^2-f^2)=H,
	\eea
	where the parameters $\lambda$, $H$ and $f$ are determined by pion decay constant $f_\pi$, sigma masses $(m_\sigma)$, and pion masses $(m_\pi)$  as,
	\begin{align}
	&\lambda =\frac{m_\sigma^2-m_{\pi}^2}{2 f_{\pi}^2},
	&&H=f_\pi m_\pi^2,
	&&&f^2= f_\pi^2 \frac{m_\sigma^2-3 m_\pi^2}{m_\sigma^2-m_\pi^2}.
	\end{align}
	For quantitative estimation, we consider decay constant $f_\pi=93$ MeV, $\sigma$ mass takes one of the values $m_\sigma=\{400, 500,700\}$, and vacuum pion mass $m_\pi=140$ MeV. We perform the analysis on the isospin pion basis which represents the physical pions. The physical pions are related to Cartesian pion fields as follows,
	\begin{align}
	&\pi^0=\pi_3,
	&&\pi^+= \frac{1}{\sqrt{2}}(\pi_1+ i \pi_2),
	&&\pi^-= \frac{1}{\sqrt{2}}(\pi_1- i \pi_2)
	\end{align}
	The interaction Lagrangian can be expressed in terms of physical pion basis as,
	\bea
	\mathcal{L}_{int}&=& \frac{\lambda}{4}\bigg(\sigma^4+(\pi^0)^4+(\pi^+)^4+(\pi^-)^4+2 (\pi^0)^2 (\pi^+)^2+2 (\pi^0)^2 (\pi^-)^2+2 (\pi^0)^2 \sigma^2\nn\\
	&+&2 (\pi^+)^2 (\pi^-)^2+2 (\pi^+)^2 \sigma^2+2 (\pi^-)^2 \sigma^2+4 v \sigma (\pi^0)^2+4 v \sigma (\pi^+)^2+4 v \sigma (\pi^-)^2+4 v \sigma^3\bigg).
	\label{Lagrangian}
	\eea
	The probable interactions in the medium can be read off from the above interaction Lagrangian. The magnetic field will further modify the Lagrangian and the interactions in the system. In the presence of a magnetic field ${\bf B}=B \hat z$, the charged pions dynamics gets affected and the four-derivative $\partial_\mu$ can be replaced with the covariant derivative $D_\mu=\partial_\mu+q A_\mu$. This modifies the kinetic part of the Lagrangian density for the charged pions. We employ $q=e$ for $\pi^\pm$ and $A^\mu=\{0,0,x B, 0\}$ in the analysis. 

\subsection*{{Interaction frequency at finite temperature} }\label{LSMT}

	The interaction frequency $\om_a$, which is the inverse of the thermal relaxation time,  for the interaction $a+b \rightarrow c+d$ can be defined as~\cite{Heffernan:2020zcf, Abhishek:2017pkp},
	\bea
	\omega^a_{\text{th}}(E_a)&\equiv&\tau_{ a}^{-1}(E_a)=\sum_{bcd}\frac{1}{1+\delta_{cd}}\int \frac{d^3 p_b d^3 p_c d^3 p_d}{(2\pi)^5} 
	\frac{|\mathcal{M}(a b\rightarrow c d)|^2}{16 E_a E_b E_c E_d}\delta^4(p_a+p_b-p_c-p_d)f_b^{0},
	\label{om}
	\eea
where $|\mathcal{M}(a b\rightarrow c d)| $ is the matrix element associated with the interaction. The particle kinematics can be further simplified in the center of mass frame (CoM). In CoM,  Eq.~\eqref{om} can be described as,    
	\bea
	\om^a_{\text{th}}&=&\frac{1}{256 \pi^3 E_a} \sum_{bcd}\frac{1}{1+\delta_{cd}}\int_{m_b}^{\infty} dE_b\sqrt{E_b^2-m_b^2} \int_{-1}^{1}\frac{dx}{p_{ab}\sqrt{s}}(t_{\text{max}}-t_{\text{min}})\,\,|\mathcal{M}|^2f_b^{0}(E_b),
	\label{om_th}
	\eea
	where the parameters take the following forms,
	\bea
	s&=& 2 E_a E_b \bigg(1+\frac{m_a^2+m_b^2}{2E_a E_b}-\frac{p_a p_b}{E_a E_b}x\bigg),\\
	t_{\text{max}}&=& m_a^2+m_c^2-\frac{1}{2s} (s+m_a^2-m_b^2)(s+m_c^2-m_d^2)+ \frac{1}{2s}\sqrt{\lambda(s,m_a^2,m_b^2)\lambda(s,m_c^2,m_d^2)},\\
	t_{\text{min}}&=& m_a^2+m_c^2-\frac{1}{2s} (s+m_a^2-m_b^2)(s+m_c^2-m_d^2)- \frac{1}{2s}\sqrt{\lambda(s,m_a^2,m_b^2)\lambda(s,m_c^2,m_d^2)},\\
	p_{ab}(s)&=&\frac{1}{2\sqrt{s}}\sqrt{\lambda(s,m_a^2,m_b^2)}.
	\eea
	The kinematic function $\lambda$ is given as $\lambda(x,y,z)=x^2+y^2+z^2-2(xy+yz+zx)$. For the case of pure thermal medium, we use the following matrix elements in Eq.~\eqref{om} from the LSM model,
	\bea
	\mathcal{M}_{fi}(\sigma\sigma|\sigma\sigma)&=&-6\lambda, \\
	\mathcal{M}_{fi}(\pi^g\pi^g|\pi^g\pi^g)&=&-6\lambda ,\,\,\{g=0,+,-\}\\
	\mathcal{M}_{fi}(\pi^+\pi^-|\pi^+\pi^-)&=&-2\lambda, \\
	\mathcal{M}_{fi}(\pi^0\pi^0|\sigma\sigma)&=&-2\lambda, \\
	\mathcal{M}_{fi}(\pi^g\sigma|\pi^g\sigma)&=&-2\lambda ,\,\,\{g=0,+,-\}\\
	\mathcal{M}_{fi}(\pi^h\pi^0|\pi^h\pi^0)&=&-2\lambda ,\,\,\{h=+,-\}.
	\eea
	Note that a pole arises in each $s, t, u$ channel while estimating the matrix element. The current analysis is on the limit with $s, t, u\rightarrow \infty$, which excludes the 3-point interactions. 
    More detailed discussions on the interaction rates and the matrix elements for the thermal medium can be found in Ref.~\cite{Heffernan:2020zcf}. In the next section, we explore the the impact of a strong magnetic field on electric charge transport and on the interaction frequency in the hadronic medium.

 \section{Longitudinal electrical conductivity in  strongly magnetized hadronic matter} \label{ES}
The dynamics of charged particles are constrained in $(1+1)-$dimensional space in the presence of a strong uniform magnetic field via Landau level quantization. We consider ${\bf B}=B\hat{z}$ in the analysis. The Landau level energy dispersion for a charged boson in a strong magnetic field can be defined as,
\begin{align}
    E_n=\sqrt{p_z^2+m^2+(2n+1) qB},
\end{align}
where $m$ is the mass of the particle and $n$ is the order of the Landau level. In the regime $T^{2}\ll \mid qB\mid$, the charged particles will be in the lowest Landau level (LLL) state, $i.e.$, $n=0$, as the  thermal occupation of higher levels is suppressed due to the Boltzmann factor  $e^{-\frac{\sqrt{qB}}{T}}$. To separate out the impact of the magnetic field on the charged particles in the medium, the equilibrium energy-momentum tensor of the system can be defined as,
\begin{align}
    T^{\mu\nu}=\bar{T}^{\mu\nu}+T_{B}^{\mu\nu},
\end{align}
where $\bar{T}^{\mu\nu}$ is the magnetic field independent part and $T_{B}^{\mu\nu}$ is the component that depends on the magnetic field in the hadronic medium. The microscopic definition and tensor decomposition of $\bar{T}^{\mu\nu}$ has the following form,
\begin{align}\label{T}
&\bar{T}^{\mu\nu}=\sum_{a=\sigma,\pi^0}\int\frac{d^3 \textbf{p}}{(2\pi)^3}{p}_a^{\mu}{p}_a^{\nu}\,f^0_a\,, 
&& \bar{T}^{\mu\nu}=\bar{\varepsilon} u^{\mu}u^{\nu}-\bar{P}\Delta^{\mu\nu},
\end{align}
where $u_\mu$ is the fluid velocity and $\Delta^{\mu\nu} = g^{\mu\nu} - u^\mu u^\nu$ is the projection operator orthogonal to $u_\mu$. Here, $\bar{\varepsilon}$ and $\bar{P}$ denotes the energy density and pressure of neutral particles. Similarly, the magnetic field dependent part of the energy-momentum tensor $T_{B}^{\mu\nu}$ at the LLL can be defined as,
\begin{align}\label{TB}
&{T_B}^{\mu\nu}=\dfrac{\mid eB\mid}{2\pi}\sum_{a=\pi^{\pm}}\int\frac{ dp_z}{2\pi}{p}_{{\|}\, a}^{\mu}{p}_{{\|}\, a}^{\nu}\,f^0_{B\, a}\,, 
&& T_B^{\mu\nu}=\varepsilon_{\|} u^{\mu}u^{\nu}-P_{\|}\Delta_{\|}^{\mu\nu},
\end{align}
with $\Delta_{\|}^{\mu\nu}\equiv g_{\|}^{\mu\nu}-u^{\mu}u^{\nu}$ is the longitudinal projection operator where $g_{\|}^{\mu\nu}=(1,0,0,-1)$. The quantities 
$P_{\|}$ and $\varepsilon_{\|}$ represent the longitudinal pressure and energy density of the charged particles in the presence of the strong magnetic field.  The charged particle motion is constrained in the direction of the magnetic field, and the Landau quantization modifies the integration phase factor and the distribution function $f^0_{B\, a}$. 
Thermodynamic quantities of the hadronic medium can be obtained by taking appropriate components of Eq.~(\ref{T}) and Eq.~(\ref{TB}). Here, we particularly focus on the specific heat capacity and speed of sound in the hadronic medium. 

With the $1+1-$dimensional motion of the charged particles in the presence of a strong magnetic field, the longitudinal current density can be defined as,
\begin{align}
    {\bf j}_{\|}=\sigma_{\|} {\bf E},
\end{align}
where $\sigma_{\|}$ is the magnetic field-dependent longitudinal conductivity in the medium. In terms of the particle distribution function, the longitudinal current density at the LLL takes the following form,
\begin{align}\label{I.35}
    {j}_{\|}=\dfrac{\mid eB\mid}{2\pi}\sum_{a=\pi^{\pm}}\int\frac{ dp_z}{2\pi}\frac{p_z}{\epsilon_{ a}}q_a\delta f_{B\, a},
\end{align}
where $\delta f_{B\, a}$ is the non-equilibrium part of the distribution function of charged pions in the strong magnetic field. As the charged particle motion is along the direction of the magnetic field, we have $({\bf v}\times {\bf B})=0$. This indicates that the Hall current will vanish in a strongly magnetized medium. Solving the Boltzmann equation in the presence of a strong magnetic field, we obtain $\delta f_{B\, a}$, and the longitudinal current can be expressed as,
\begin{align}
    {\bf j}_{\|}=\dfrac{\mid eB\mid}{2\pi}\sum_{a=\pi^{\pm}}\int\frac{ dp_z}{2\pi}\frac{p^2_z}{\epsilon^2_{ a}}q_a^2\tau_{B\,a}\Big(-\frac{\partial f_{B\,a}^0}{\partial\epsilon_a}\Big)   {\bf E}.
\end{align}
Here, $\tau_{B\,a}$ denotes the magnetic field-dependent thermal relaxation in the strongly magnetized hadronic medium. Unlike in the case of a weakly magnetized medium, particle interaction frequency will depend upon the strength of the magnetic field in the strong field regime. The same observation holds true for the case of the QGP medium~\cite{Hattori:2016lqx}. Now, we proceed with the discussion of thermal relaxation time in a strongly magnetized hadronic medium at finite temperature.

\subsection*{{Magnetic field-dependent interaction frequency at finite temperature} }\label{LSMB}

In the presence of a strong magnetic field, the matrix amplitudes, as well as the interaction rates, are modified as the charged particles interact with the magnetic field~\cite{Ghosh:2022xtv}. In the strong magnetic field limit, LLL approximation is considered in the calculations. For $\pi^+(k_a)+\pi^+(k_b)\rightarrow \pi^+(k_c)+\pi^+(k_d)$ and $\pi^-(k_a)+\pi^-(k_b)\rightarrow \pi^-(k_c)+\pi^-(k_d)$ processes, the interaction frequency of $\pi^h(h=\pm)$ can be defined as follows,
	\bea
	\omega^{a}_1 &=&\frac{1}{2} \int \frac{dk_y^b dk_z^b}{(2\pi)^2}\frac{dk_y^c dk_z^c}{(2\pi)^2}\frac{dk_y^d dk_z^d}{(2\pi)^2}
	(2\pi)^3\delta^{(3)}_{\tilde x}(k_a+k_b-k_c-k_d)\nonumber\\
	&\times&\frac{1}{16 E_a E_b E_c E_d}|\mathcal{M}_{fi}(\pi^\pm \pi^\pm|\pi^\pm\pi^\pm)|^2 f_b^{0}\label{matrix_el},
	\eea
	where $\delta^{(3)}_{\tilde x}$ is the $\delta-$function for space-time coordinates except $x$. The matrix element is evaluated from the S-matrix calculations by using the solutions of the Klein-Gordon equation for charged particle in the presence of a strong magnetic field and  takes the form as,
	\bea
	|\mathcal{M}_{fi}|^2&=&(6\lambda)^2 \frac{|eB|}{2\pi}\exp\left\{-\frac{(k_y^a+k_y^b+k_y^c+k_y^d)^2-4(k_y^a)^2-4(k_y^b)^2-4(k_y^c)^2-4(k_y^d)^2}{4|eB|}\right\}.
	\label{matrix_elem}
	\eea
	After performing the integration over $k_y^b$, $k_y^c$, $k_y^d$, and using the properties of $\delta$-function, Eq.~(\ref{matrix_el}) can be simplified as follows,
	\bea
	\omega^a_1&=& 
     9{\lambda^2}\frac{|eB|^2}{(4\pi)^3}\int_{-\infty}^{\infty} dk_z^b\, \frac{1}{E_a^2 E_b^2}\bigg(\frac{k_z^a}{E_a}-\frac{k_z^b}{E_b}\bigg)^{-1}f_b^{0}.
	\eea
	Similarly, the interaction rate of $\pi^{h}(h=\pm)$ for the scattering process
	$\pi^+(k_a)+\pi^-(k_b)\rightarrow \pi^+(k_c)+\pi^-(k_d)$  can be described as,
	\bea
	\om^a_2&=&2\lambda^2\frac{|eB|^2}{(4\pi)^3}\int_{-\infty}^{\infty} dk_z^b\, \frac{1}{E_a^2 E_b^2}\bigg(\frac{k_z^a}{E_a}-\frac{k_z^b}{E_b}\bigg)^{-1}f_b^{0}.
	\eea
	For the other scattering processes,  $\pi^h(p)+\sigma(k) \rightarrow \pi^h(p')+\sigma(k')$ 
	and $\pi^h(p)+\pi^0(k) \rightarrow \pi^h(p')+\pi^0(k')$, the interaction frequency of $\pi^h$ particle is given by
	\bea
	\om^{\pi^h}_3(p)&=& \int\frac{d^3k}{(2\pi)^3}\frac{d^3k'}{(2\pi)^3}\frac{d^2p'}{(2\pi)^2}
	\frac{(2\pi)^3 \delta^{(3)}_{\tilde x}(p+k-p'-k')}{16E_kE_pE_{p'}E_{k'}}\nn\\
	&\times&(2\lambda)^2 \exp\left\{-\frac{(p_y-p_y')^2+(k_x-k_x')^2}{2|eB|}\right\} f^{0}_{\sigma/\pi_0} (E_k).
	\label{matrix_e}
	\eea
	Following the similar procedure as that for other processes, Eq.~(\ref{matrix_e}) can further simplified by performing the integration over $p_y'$ and $k_z$ and has the following form, 
	\bea
	\om^{\pi^h}_3&=&
	\Big(\frac{\lambda}{2}\Big)^2 \int\frac{d^3k'}{(2\pi)^5}\frac{d^2 k_\perp dp_z'\,\,\, 2|E_{p'}+E_{k'}-E_p|}{E_pE_k'E_{p'}\sqrt{k_\perp^2+(p_z'+k_z'-p_z)^2}}
	\exp{\left\{-\frac{(k-k')_\perp^2}{2|eB|}\right\}}\nn\\
	&\times&\delta(k^2_\perp-(E_{p'}+E_{k'}-E_p)^2+(p_z'+k_z'-p_z)^2)f_{\sigma/\pi_0}^{0}\bigg(\sqrt{k_\perp^2+(p_z'+k_z'-p_z)^2}\bigg).
	\eea
	After the $k_\perp $ integration, we perform other integration numerically for the quantitative estimation.
	
	To calculate the frequency of interaction of the neutral scalar particles $\sigma$ and $\pi^0$, the processes $\pi^h(p)+\sigma(k) \rightarrow \pi^h(p')+\sigma(k')$ 
	and $\pi^h(p)+\pi^0(k) \rightarrow \pi^h(p')+\pi^0(k')$ need to be considered. The matrix element of there interactions takes the same form as described in Eq.~\eqref{matrix_elem}.  After performing the integration over $k_z'$ and $p_y'$, the expressions of the interaction frequencies of $\pi^0$ and $\sigma$ read as,
	\bea
	\om^{\sigma,\pi^0}_4&=&\Big(\frac{\lambda}{2}\Big)^2\int\frac{d^2k'_\perp}{(2\pi)^4 L_x}\frac{d p_y dp_z\, dp_z'}{E_pE_{k}E_{p'}\sqrt{k_{\perp}'^2+(p_z+k_z-p_z')^2}}\nn\\
	&\times&\delta(E_p+E_k-\sqrt{k_{\perp}'^2+(p_z+k_z-p_z')^2}-E_{p'})\exp{\left\{-\frac{(k-k')_\perp^2}{2|eB|}\right\}}f_{\pi^h}^{0}(E_p).
	\eea
	Considering the infinite volume limit of the finite box side $L_x$ and using the properties of $\delta-$ function, the interaction can be further evaluated numerically to obtain the thermal relaxation time associated with the process.  
	
	Now, we briefly discuss the procedure to evaluate the total relaxation time, say the relaxation time of $\pi^+$, $i.e.$,  $\tau_{B\, \pi^+}$ in the medium. 
    To that end, we need to consider the following processes, 
	\bea
	&&\pi^++\pi^g\rightarrow \pi^++\pi^g \,\,\,(g=+,-,0),\nn\\
	&&\pi^++\sigma\rightarrow \pi^++\sigma. 
	\eea
	The total interaction frequency for $\pi^{+}$ can be expressed as $\om^{\pi^+}=\om_{1}^{\pi^+}+\om_{2}^{\pi^+}+\om_{3}^{\pi^+}$. In the similar fashion we can calculate the interaction frequencies for other particles.  As the charged neutral particles are not directly affected by the magnetic field,  interaction rate associated with the scalar particle interaction, for example, $\sigma\sigma \rightarrow \sigma\sigma$, remain intact as described in Eq.~\eqref{om_th}. We obtain the longitudinal conductivity in a strongly magnetized hadronic medium by employing these magnetic field-dependent interaction rates. 

	\section{Hadronic medium response to a time-varying electric field}\label{NE}
Here, the effects of the time dependence of the external fields are considered on the electric charge transport in the hadronic medium. The present focus is on the case in which the time inhomogeneity of the external field is small such that the collisional properties of the medium are not negligible. First, we consider the case of a weakly magnetized hadronic matter with ${\bf E}\cdot{\bf B}=0$.  The vector quantities that can act as the source of the induced current are $\textbf{E}, (\textbf{E}\times \dot{\textbf{B}}), \textbf{B}, \dot{\textbf{E}}, \dot{\textbf{B}}, (\dot{\textbf{E}}\times \textbf{B}), (\textbf{E}\times \dot{\textbf{B}})$. These quantities can be further related by Maxwell's equations. Notably, the parity of the current density operator is different from that of ${\bf B}$ and $\dot{\bf B}$. This indicates that the components of the current density associated with ${\bf B}$ and $\dot{\bf B}$ cannot exist due to the parity considerations and the choice of direction of electromagnetic fields. The first step toward the estimation of various components of current density is to obtain the non-equilibrium part of the distribution function. For a weakly magnetized hadronic medium, $\delta f_{a}$ in the presence of a time-varying electric field can be described as follows,
\begin{align}\label{I.47}
\delta f_k= {\bf{p}}\cdot\Big[\alpha_1\textbf{E} + \alpha_3(\textbf{E}\times \textbf{B})+\beta_1\dot{\textbf{E}}+\beta_3(\dot{\textbf{E}}\times \textbf{B})\Big] \frac{\partial f^0_k}{\partial \epsilon}. 
\end{align}
Here, $\beta_1$ and $\beta_3$ denote the additional coefficients that give rise to the components of current density due to the time evolution of the electric field in the hadronic medium. The coefficients can be obtained by solving the Boltzmann equation for the case of time-varying fields.  By employing Eq.~(\ref{I.47}) in Eq.~(\ref{I.3}), we have
\begin{align}\label{I.48}
& \epsilon_a {\bf v}_a\cdot\Big[\alpha_1 \dot{{\bf E}}+\alpha_3 (\dot{{\bf E}}\times {\bf B}) +\mathcal{O}(\ddot{\bf E})\Big]
 +q_{a}{\bf v}_a\cdot{\bf E}-\alpha_1 q_{a}{{\bf v}_a}\cdot({\bf E}\times {\bf B})-\beta_1 q_{a}{\bf v}_a\cdot(\dot{{\bf E}}\times {\bf B})\nonumber\\
& +\alpha_3 q_{a}({\bf v}_a\cdot{\bf E})(B^2)-\alpha_3 q_{a}({\bf v}_a\cdot{\bf B})({\bf B}\cdot{\bf E})+\beta_3 q_{a}({\bf v}_a\cdot\dot{{\bf E}})(B^2)-\beta_3 q_{a}({\bf v}_a\cdot{\bf B})({\bf B}.\dot{{\bf E}})\nonumber\\
& =-\frac{\epsilon_a}{\tau_a}\Big[\alpha_1 {\bf v}_a\cdot{\bf E}+\beta_a {\bf v}_a\cdot\dot{{\bf E}}+\alpha_3 {\bf v}_a\cdot({\bf E}\times {\bf B})+\beta_3 {\bf v}_a\cdot(\dot{{\bf E}}\times {\bf B})\Big].
\end{align}
Note that the terms with higher-order derivatives are neglected in the analysis. Comparing various tensorial structures on both sides of Eq.~(\ref{I.48}) and solving the obtained coupled equations, we estimate the transport coefficients associated with electric charge transport of the weakly magnetized hadronic medium in the presence of a time-evolving electric field. We observe that the forms of coefficients $\alpha_1$ and $\alpha_3$ (as described in Eq.~(\ref{I.7})) remain intact due to the time dependence of the electric field. However, we have obtained an additional set of coupled equations while comparing terms with $ ({\bf v}_a\cdot\dot{\bf E})$ and ${\bf v}_a.(\dot{{\bf E}}\times {\bf B})$ in both sides of Eq.~(\ref{I.48}) as,
\begin{align}\label{I.49}
& -\frac{\epsilon_a}{\tau_{a}}\beta_1=\epsilon_a\alpha_1+\beta_3q_{a} B^2,
&& -\frac{\epsilon_a}{\tau_{a}}\beta_3=\epsilon_a\alpha_3-\beta_1q_{a}.
\end{align}
Employing the form of $\alpha_1$ and $\alpha_3$, the Eq.~(\ref{I.49}) can be solved and the coefficient $\beta_1$ and  $\beta_1$ take the following forms,
\begin{align}
&\beta_1=\frac{ \epsilon_a q_a\Big[\frac{\epsilon^2_a}{\tau^2_a}-(q_aB)^2\Big]}{\Big[\frac{\epsilon^2_a}{\tau^2_a}+(q_aB)^2\Big]^2},
&& \beta_2=\frac{ 2\epsilon^2_a q^2_a}{\tau_a\Big[\frac{\epsilon^2_a}{\tau^2_a}+(q_aB)^2\Big]^2}.
\end{align}
Hence, the current density in a weakly magnetized hadronic matter due to the time-evolving electric field can be defined as,
\begin{align}\label{I.51}
{\bf j}=&\frac{1}{3}\sum_a \int \frac{d^3 \textbf{p}}{(2\pi)^3} \frac{p^2}{\epsilon_a}q_a\bigg\{\frac{\epsilon_a q_a\tau_a}{(\tau_{a}^2 q_a^2B^2+\epsilon_a^2)}{\bf E}+\frac{ q^2_a\tau^2_a}{(\tau_{a}^2 q_a^2B^2+\epsilon_a^2)}({\bf E}\times{\bf B})-
\frac{ \epsilon_a q_a\Big[\frac{\epsilon^2_a}{\tau^2_a}-(q_aB)^2\Big]}{\Big[\frac{\epsilon^2_a}{\tau^2_a}+(q_aB)^2\Big]^2}\dot{\bf E}\nonumber\\&
-\frac{ 2\epsilon^2_a q^2_a}{\tau_a\Big[\frac{\epsilon^2_a}{\tau^2_a}+(q_aB)^2\Big]^2}(\dot{\bf E}\times{\bf B})\bigg\}\Big(-\frac{\partial f_a^0}{\partial\epsilon_a}\Big).
\end{align}
In the limit of constant electromagnetic fields, Eq.~(\ref{I.51}) reduces back to Eq.~(\ref{I.8}). Note that in a system with non-zero chiral chemical potential, there will be more components to the current density which is proportional to the magnetic field in the medium. This aspect is beyond the scope of the present study.

For the case of a strongly magnetized hadronic medium, the component of current density due to the term ${\bf E}\times{\bf B}$ will not survive due to the $1+1-$dimensional Landau level motion of the charged particle in the medium. Solving the $1+1-$dimensional Boltzmann equation within the relaxation time approximation in the presence of a time-varying electric field, we obtain the non-equilibrium correction of the charged hadronic particle as follows,
\begin{align}\label{I.52}
\delta f_{B\, k} =\tau_{B\,a}q_a\frac{p_z}{\epsilon_{ a}}\Big(-\frac{\partial f_{B\,a}^0}{\partial\epsilon_a}\Big)   {\bf E}-\tau^2_{B\,a}q_a\frac{p_z}{\epsilon_{ a}}\Big(-\frac{\partial f_{B\,a}^0}{\partial\epsilon_a}\Big)   \dot{\bf E}.
\end{align}
Substituting Eq.~(\ref{I.52}) in Eq.~(\ref{I.35}), we obtain the current density in a strongly magnetized hadronic matter in the presence of a time-varying electric field as,
\begin{align}\label{I.53}
 {\bf j}_{\|}=\dfrac{\mid eB\mid}{2\pi}\sum_{a=\pi^{\pm}}\int\frac{ dp_z}{2\pi}\frac{p^2_z}{\epsilon^2_{ a}}q_a^2\tau_{B\,a}\Big(-\frac{\partial f_{B\,a}^0}{\partial\epsilon_a}\Big)   {\bf E}-\dfrac{\mid eB\mid}{2\pi}\sum_{a=\pi^{\pm}}\int\frac{ dp_z}{2\pi}\frac{p^2_z}{\epsilon^2_{ a}}q_a^2\tau^2_{B\,a}\Big(-\frac{\partial f_{B\,a}^0}{\partial\epsilon_a}\Big)   \dot{\bf E}.
\end{align}
Here, the first term is leading order in relaxation time and gives rise to the dominant longitudinal current density. The second term, which is higher-order in $\tau_{B\,a}$  denotes the correction due to the time dependence of the external electric field.


\section{Results and discussions }\label{RE}
\begin{figure}[tbh]
\begin{center}
\includegraphics[scale=0.62]{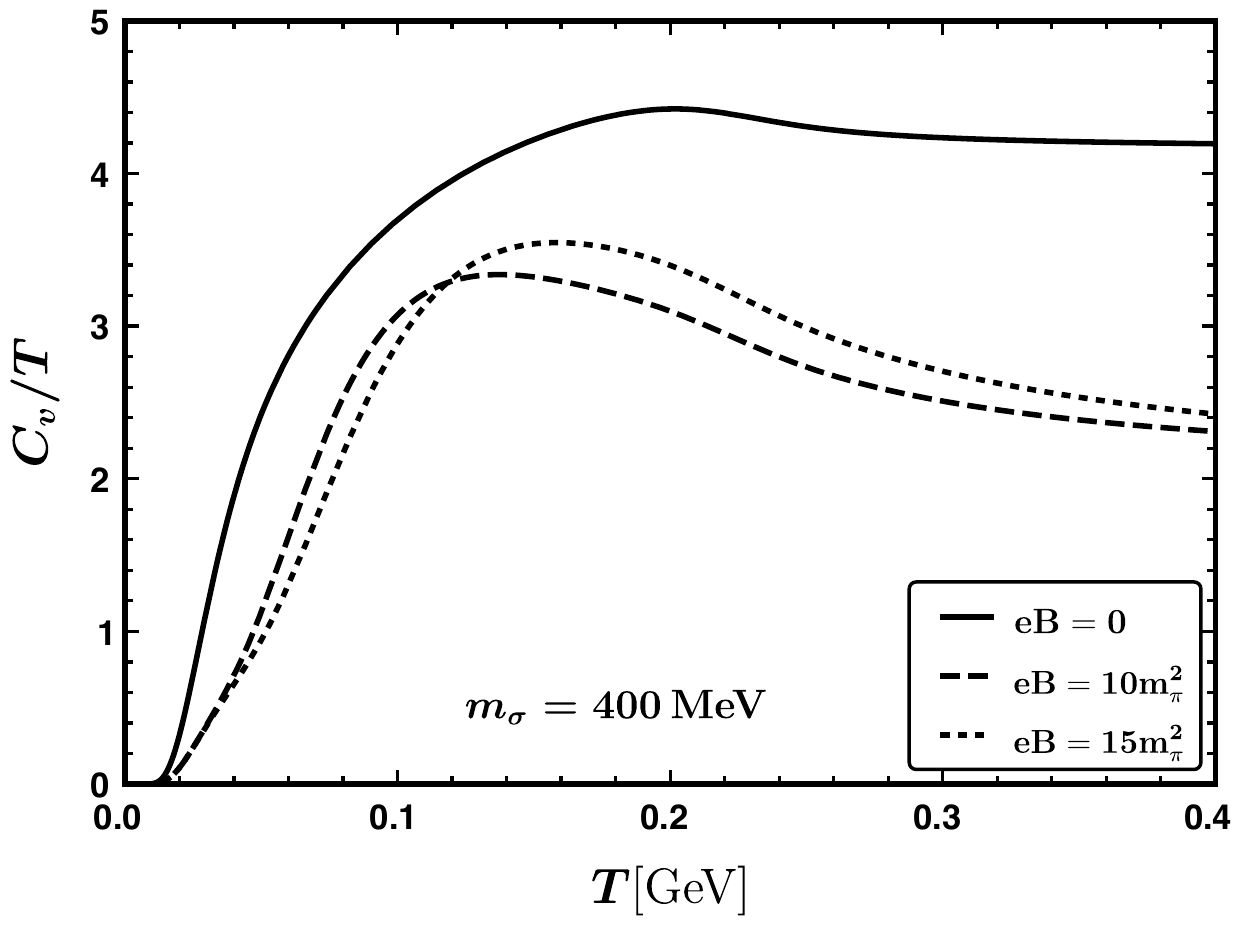}
\includegraphics[scale=0.5]{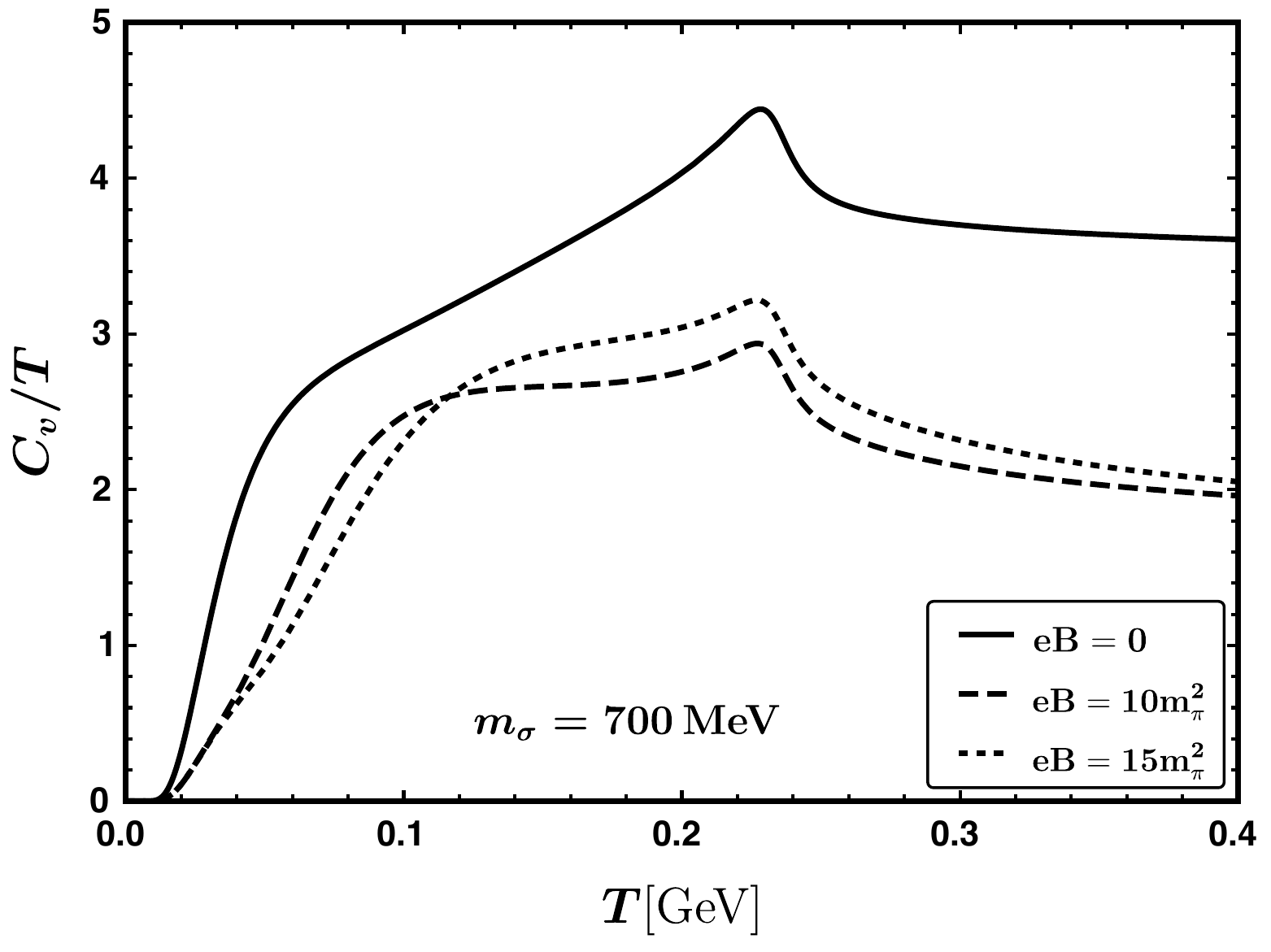}
\caption{\small Temperature behavior of $\frac{C_v}{T}$ in a magnetized hadronic matter for a vacuum sigma mass of 400 MeV (left panel) and 700 MeV (right panel). }
\label{f1}
\end{center}
\end{figure}
\begin{figure}[tbh]
\begin{center}
\includegraphics[scale=0.62]{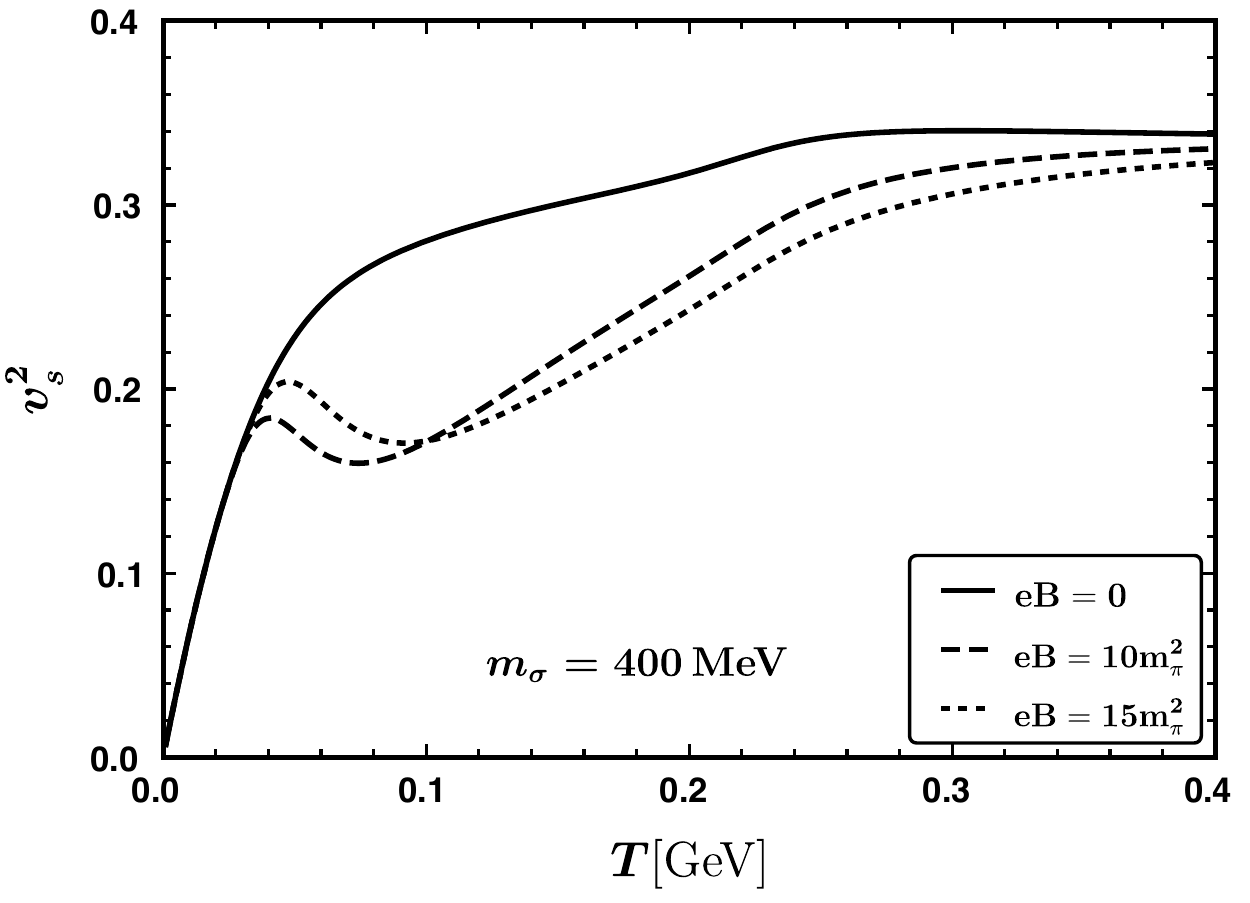}
\includegraphics[scale=0.5]{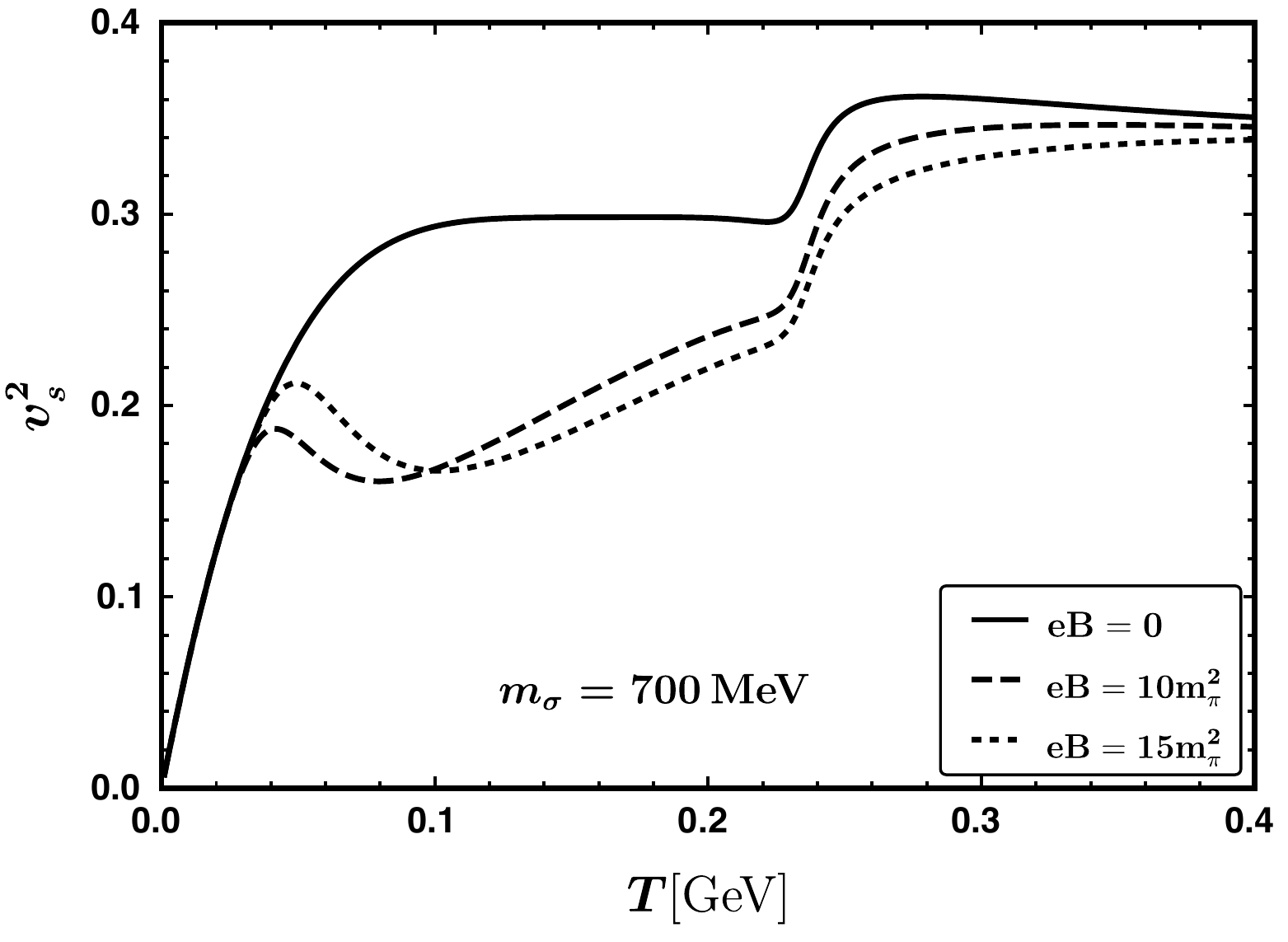}
\caption{Impact of magnetic field on the temperature behaviour of speed of sound in the hadronic medium for $m_\sigma=400$ MeV (left panel) and $m_\sigma=700$ MeV (right panel). }
\label{f2}
\end{center}
\end{figure}
\begin{figure}
\begin{center}
\includegraphics[scale=0.5]{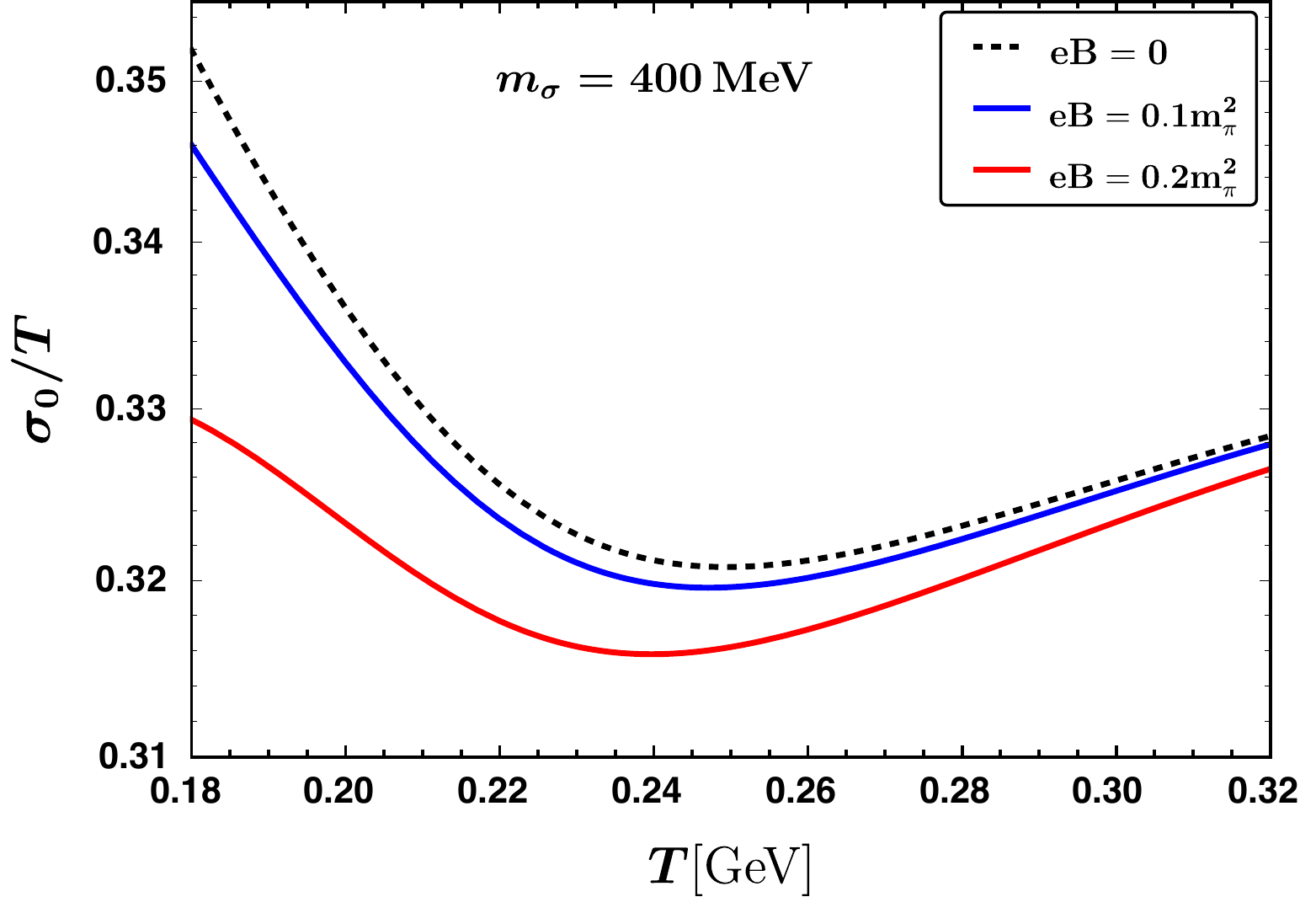}
\includegraphics[scale=0.5]{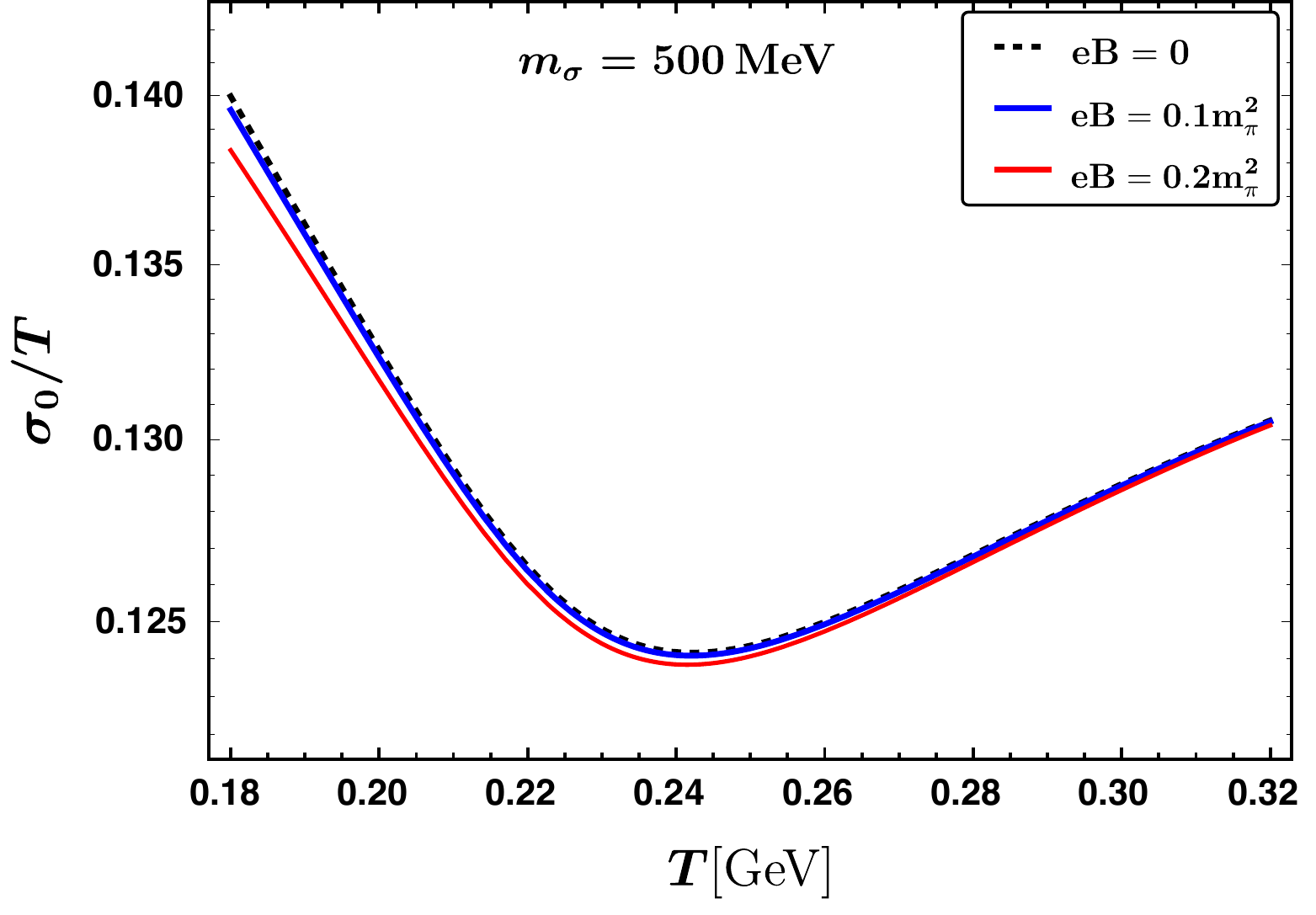}
\caption{\small Temperature dependence of $\frac{\sigma_0}{T}$ of a weakly magnetized hadronic medium for $m_\sigma=400$ MeV (left panel) and $m_\sigma=500$ MeV (right panel).}
\label{f3}
\end{center}
\end{figure}
\begin{figure}
\begin{center}
\includegraphics[scale=0.5]{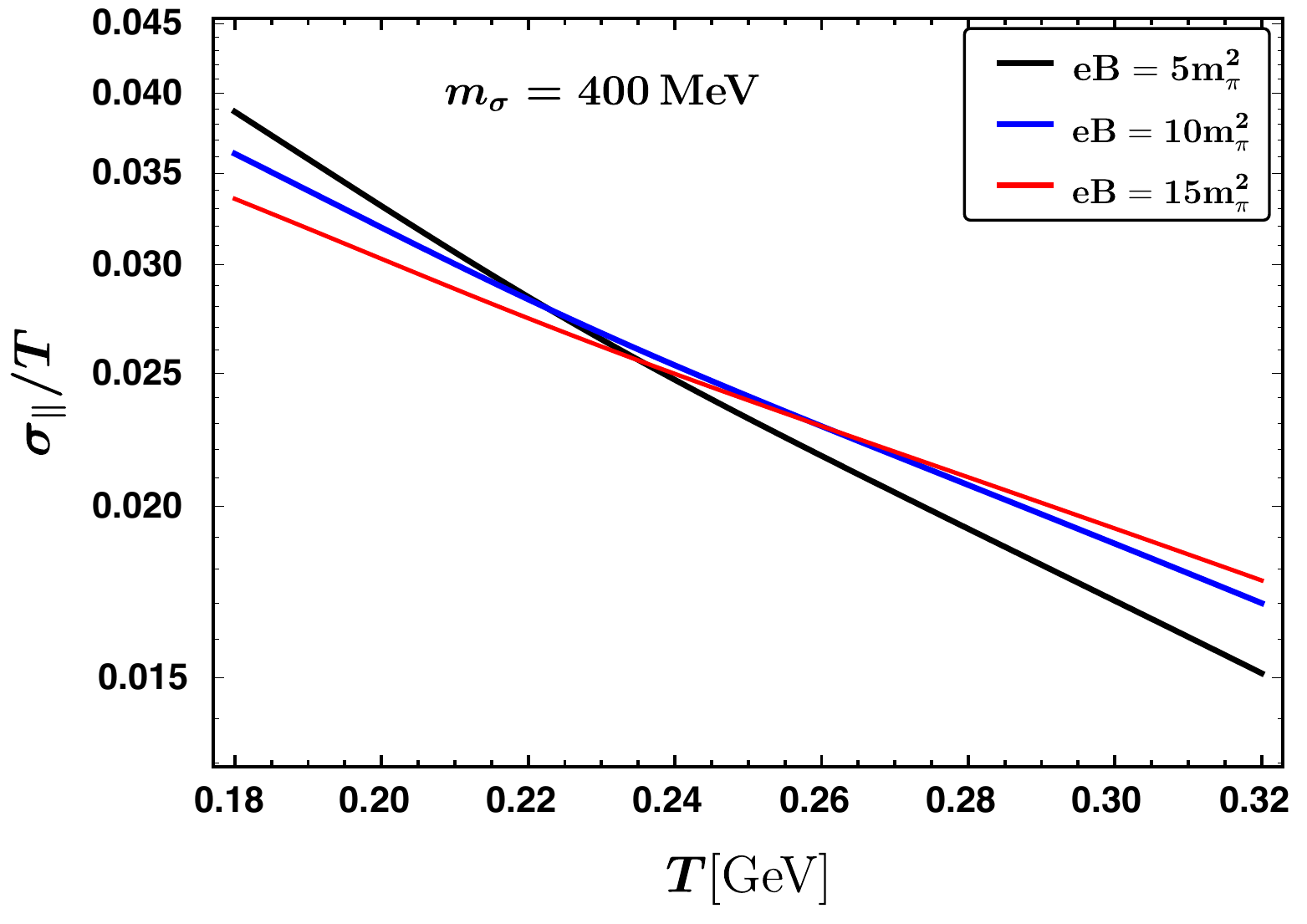}
\includegraphics[scale=0.525]{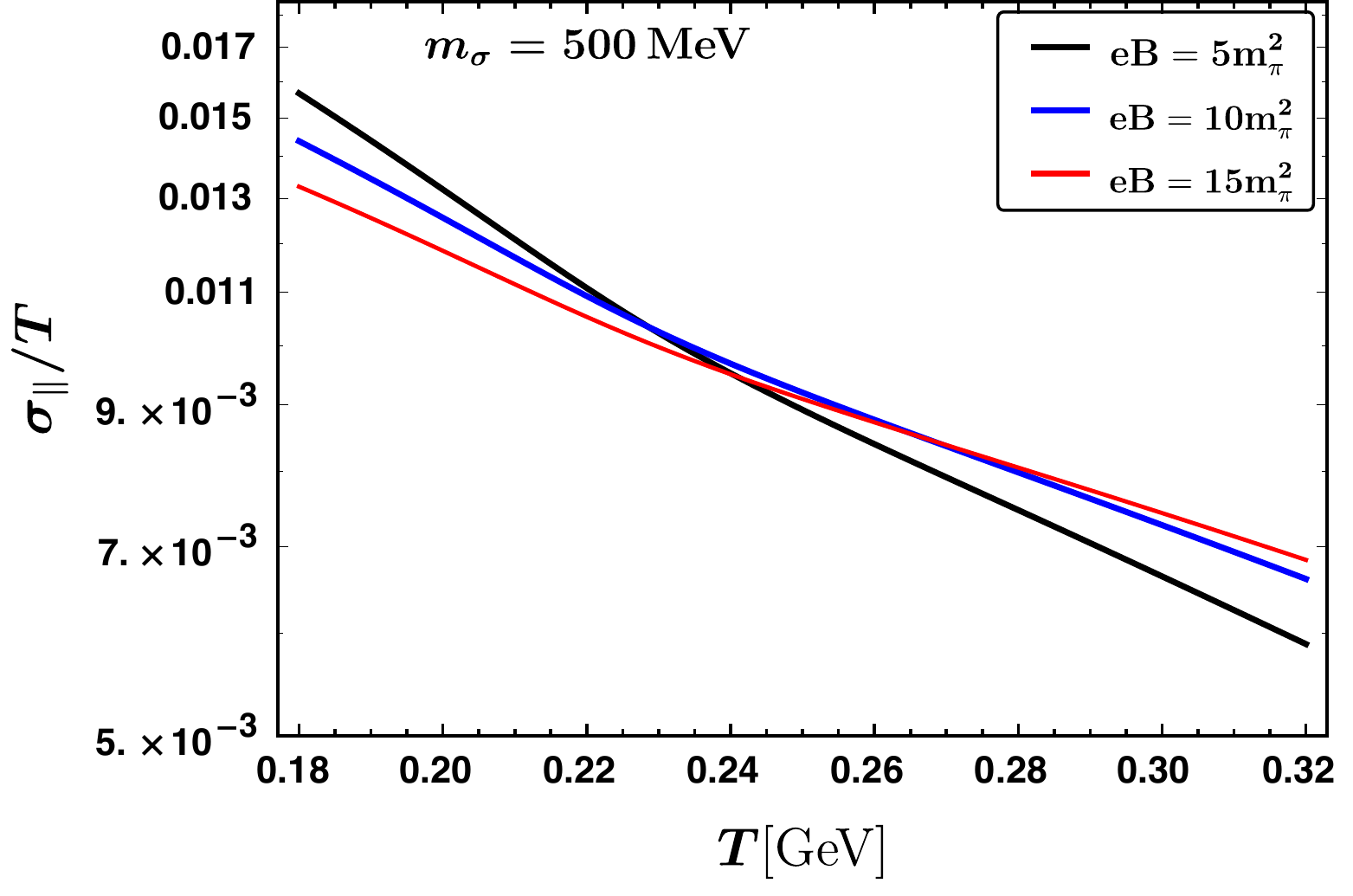}
\caption{Longitudinal conductivity of a strongly magnetized hadronic medium as a function of temperature for $m_\sigma=400$ MeV (left panel) and $m_\sigma=500$ MeV (right panel). }
\label{f4}
\end{center}
\end{figure}
We initiate the discussion with the effect of magnetic field and sigma mass on the specific heat capacity $(C_v)$ and speed of sound $(v_s)$ in the hadronic medium. These thermodynamic quantities can be described as,
\begin{align}
&C_v=\Bigg(\frac{\partial\epsilon}{\partial T}\Bigg)_v, && v_s^2=\frac{\partial P}{\partial\epsilon},
\end{align}
where $\epsilon$ and $P$ are the energy density and pressure of the hadronic matter. In the current analysis, we have neglected the effect of the magnetic field on the thermodynamics of a weakly magnetized hadronic medium as the field enters as a small perturbation in the system. However, the impact of the magnetic field has been incorporated in the strongly magnetized regime.
In Fig.~\ref{f1}, the ratio $C_v/T$ is plotted as a function of temperature for different values of sigma mass and magnetic field. It is observed that $C_v/T$ has a strong dependence on the Landau level dynamics of the charged hadronic particles in the presence of a strong magnetic field throughout the chosen temperature regime. The impact of sigma mass on the chiral symmetry restoration is studied in terms of thermodynamic quantities. We have observed a peak in the heat capacity for the case with $m_\sigma=700$ MeV, which indicates that the chiral symmetry is restored faster for higher sigma mass. The mean field effects are observed to have a significant role in the low-temperature behavior of the medium. This is reflected in the temperature behavior of speed of sound in the hadronic matter as plotted in Fig.~\ref{f2}. In contrast to the behavior of $C_v/T$ in a magnetized hadronic medium, the effect of the magnetic field on the speed of sound is more pronounced in the intermediate temperature regime ($0.05 $ to $0.3$ GeV). The speed of sound measures the conformality of the medium, and at very high temperature, $v^2_s$ reaches the Boltzmann limit with $\epsilon\approx 3P$. The magnetic field-dependent speed of sound plays a vital role in the understanding of the behavior of bulk viscosity in the magnetized hadronic medium.

The electrical conductivity $\sigma_0$ of hadronic matter in the presence of a weak magnetic field is depicted in Fig.~\ref{f3} for different choices of sigma mass. For the quantitative estimation, we consider the case of ${\bf E}\cdot{\bf B}=0$. Notably, the vacuum sigma mass has a visible impact on the conductivity of the hadronic medium. As the value of $m_\sigma$ increases, the ratio $\sigma_0/T$ decreases significantly. A detailed comparative analysis of the LSM description of electrical conductivity with other approaches at the case of vanishing magnetic field is presented in Ref.~\cite{Heffernan:2020zcf}.  In a weakly magnetized medium, the magnetic field effects enter through the cyclotron motion via the Lorentz force term in the Boltzmann equation. The impact of the magnetic field on the temperature behavior of $\sigma_0/T$ is more pronounced in the lower temperature regimes, especially for the lower value of $m_\sigma$. This is attributed due to the fact that the magnetic field act as a perturbation in a weakly magnetized hadronic medium, and at high-temperature regimes, the strength of the magnetic field becomes negligible in comparison with the temperature scale of the medium. It is important to emphasize that the Hall conductivity $\sigma_1$ as described in Eq.~(\ref{I.10}) vanishes in the present analysis as the focus is on the limit $\mu=0$.

\begin{figure}
\begin{center}
\includegraphics[scale=0.62]{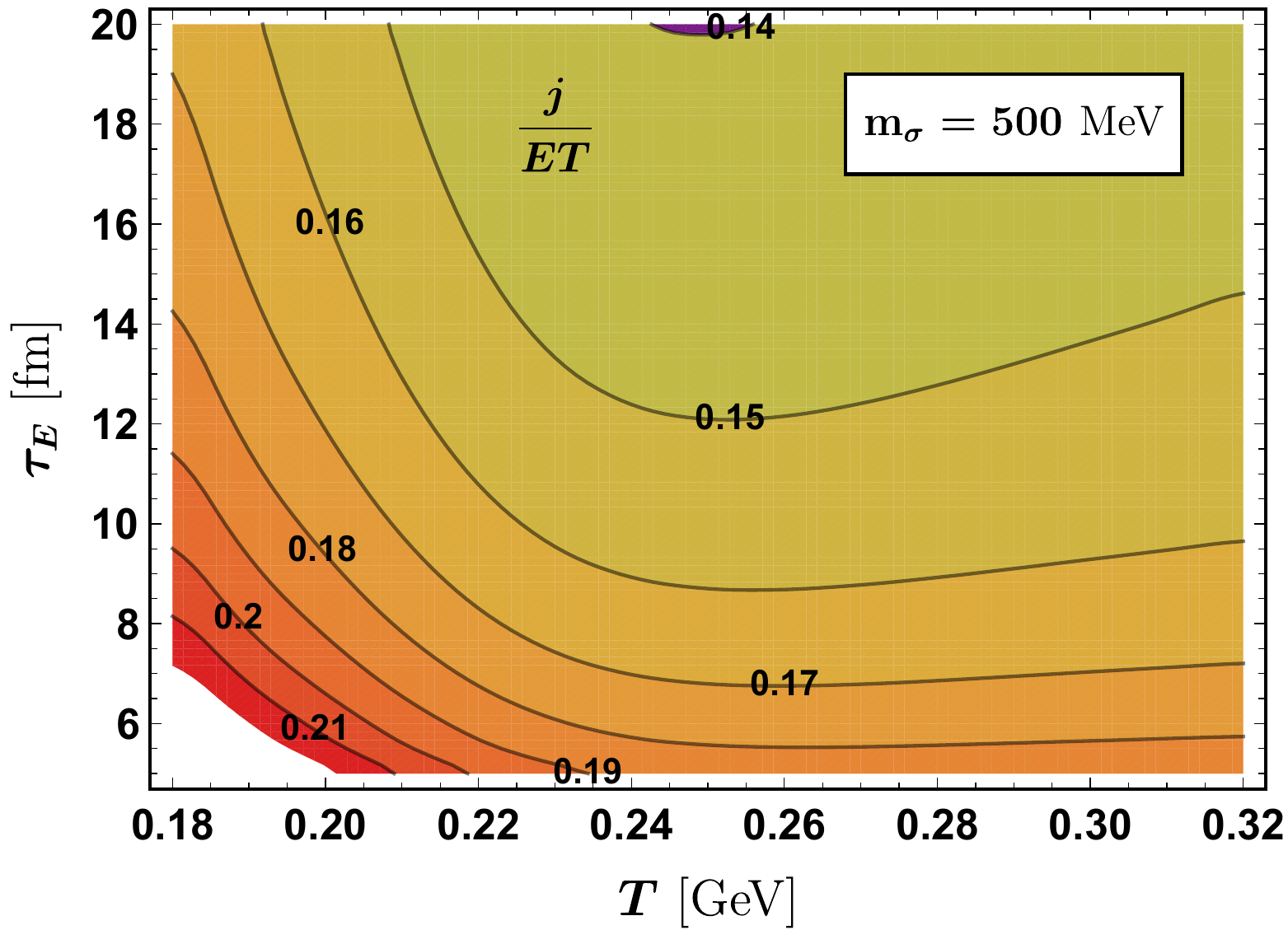}
\caption{ The effect of temperature and electric field decay parameter on $\frac{j}{ET}$ at vanishing magnetic field. Curved lines represent constant value contours of $\frac{j}{ET}$.}
\label{f4.1}
\end{center}
\end{figure}

The electric charge transport in the hadronic matter in the presence of a strong magnetic field can be quantified in terms of longitudinal current density.  In contrast to the case of a weakly magnetized medium, in a strongly magnetized hadronic medium, the magnetic field effects enter through the Landau level dispersion and through the interaction frequency. In Fig.~\ref{f4}, the longitudinal conductivity is plotted as a function of temperature for different values of magnetic field and sigma mass. It is seen that the
conductivity in a strongly magnetized hadronic medium significantly varies due to the constrained Landau level dynamics of the charged particles. The strength of the magnetic field in the medium has a visible impact on the behavior of $\sigma_{\parallel}/T$. In the low-temperature regime, $\sigma_{\parallel}/T$ decreases with an increase in the field strength. However, the behavior is quite the opposite in the high-temperature regime. The observation holds true for both the cases with $m_\sigma=400$ MeV and $m_\sigma=500$ MeV. It is important to emphasize that at higher temperatures, higher Landau level contributions may not be negligible and can affect the behavior of longitudinal transport in the medium, which is beyond the scope of the present study. Similar to the behavior of $\sigma_{0}/T$ in a weakly magnetized medium, $\sigma_{\parallel}/T$  has a strong dependence on the sigma mass. 

\begin{figure}
\begin{center}
\includegraphics[scale=0.5]{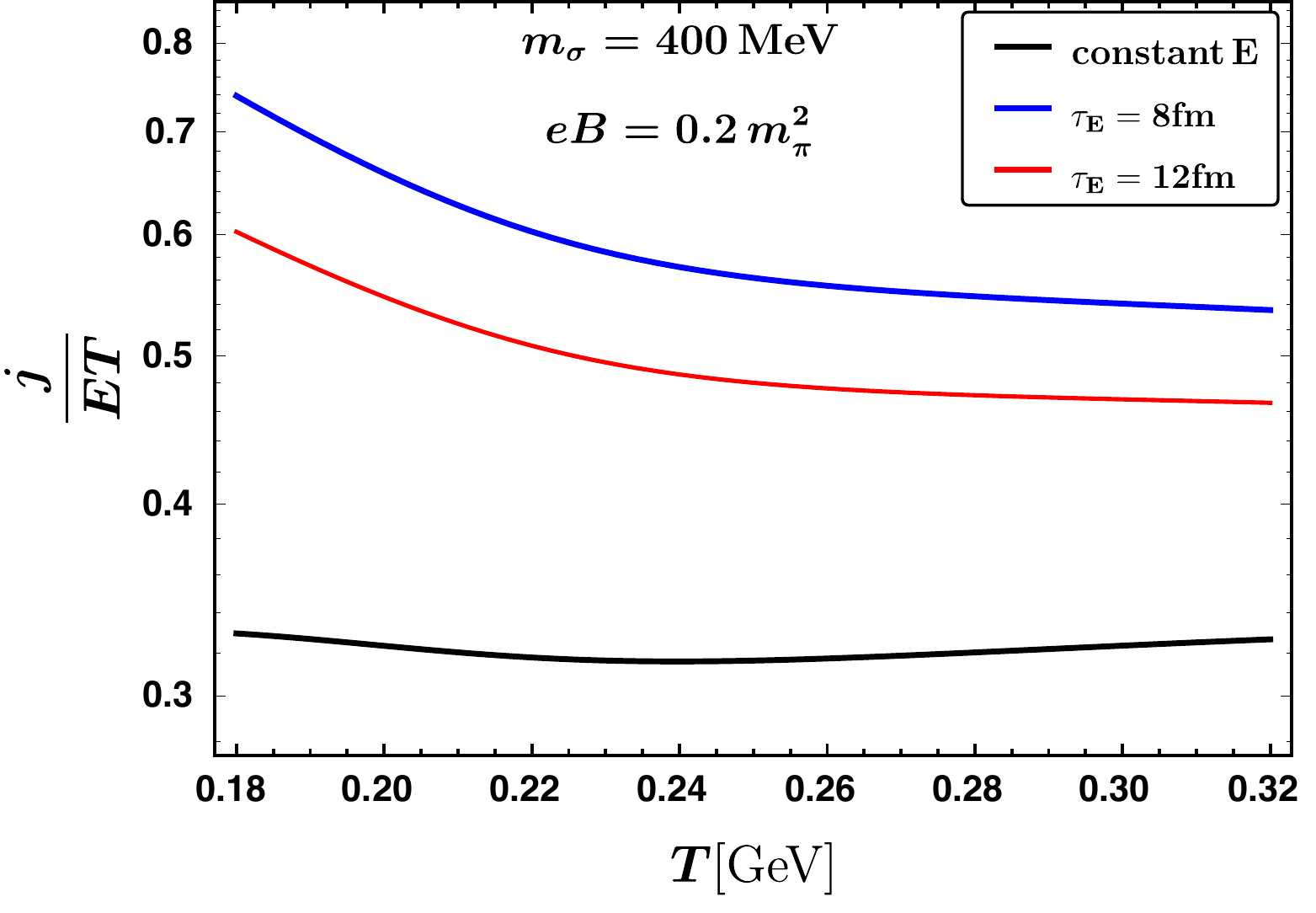}
\includegraphics[scale=0.508]{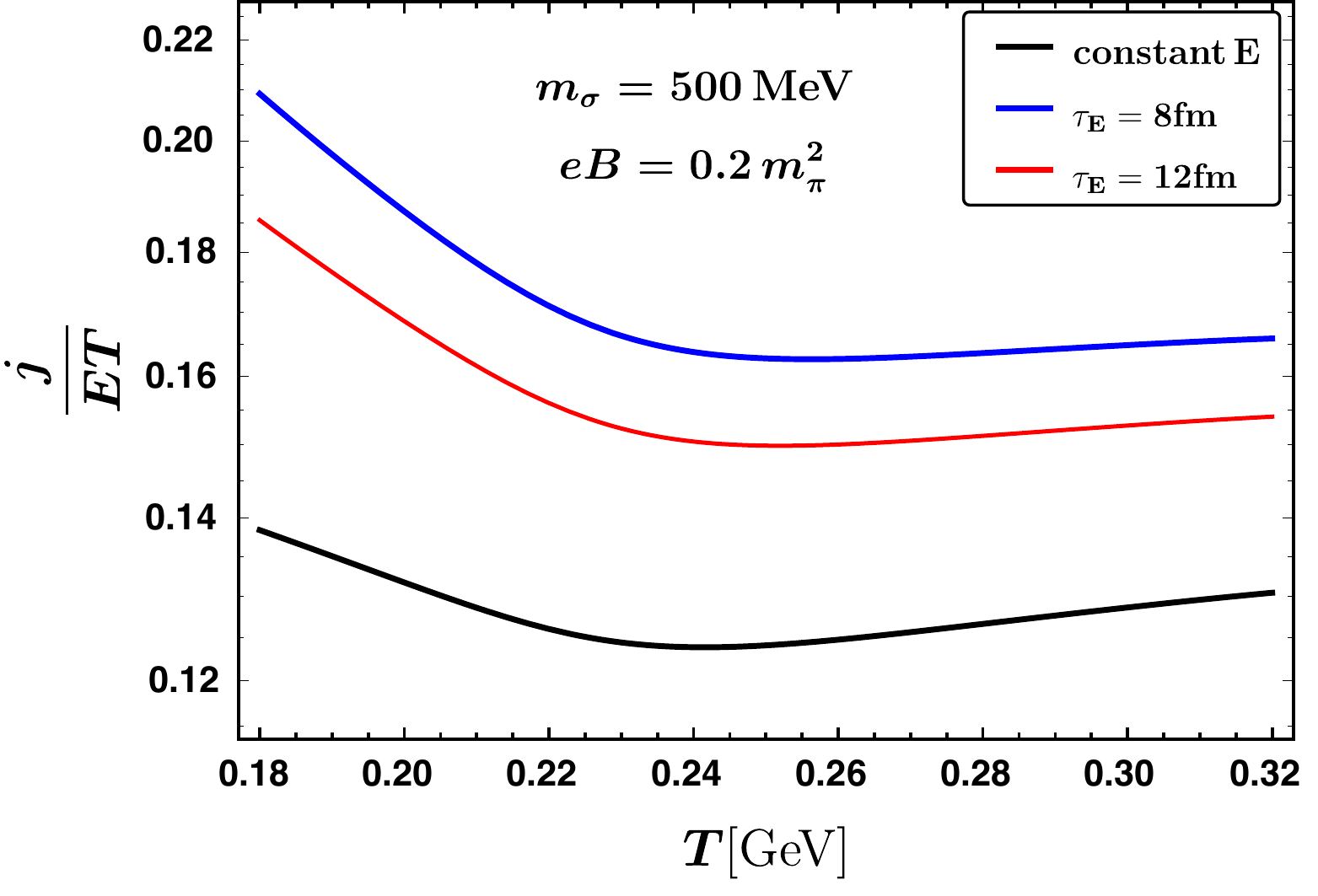}
\caption{\small Temperature dependence of $\frac{j}{ET}$  due to time-varying electric field in a weakly magnetized hadronic matter for different values of decay parameter.}
\label{f5}
\end{center}
\end{figure}

Hadronic medium response to an external time-dependent electric field is studied in terms of the ratio $\frac{j}{ET}$. In the limit of constant fields, $\frac{j}{ET}=\frac{\sigma_0}{T}$. We consider the case of a slowly varying field to include collisional aspects of the medium, and hence the induced magnetic field due to the time dependence of the electric field is neglected in the analysis. We observe that the proper time dependence of the external electric field gives rise to additional components of the current density and gives corrections to the Ohmic current in the medium. {For the quantitative estimation, we consider the exponential time decay of the electric field as described in~\cite{Hongo:2013cqa,Satow:2014lia} with $\frac{\dot{\bf E}}{\bf E}\propto -\frac{1}{\tau_E}$ where $\tau_E$ as the decay parameter.} The ratio $\frac{j}{ET}$ is plotted as a function of temperature and decay parameter of the external electric field at $eB=0$ case in Fig.~\ref{f4.1}. Time evolution of the electric field is seen to have a noticeable effect on the temperature behavior of induced current density in the hadronic medium.

\begin{figure}
\begin{center}
\includegraphics[scale=0.48]{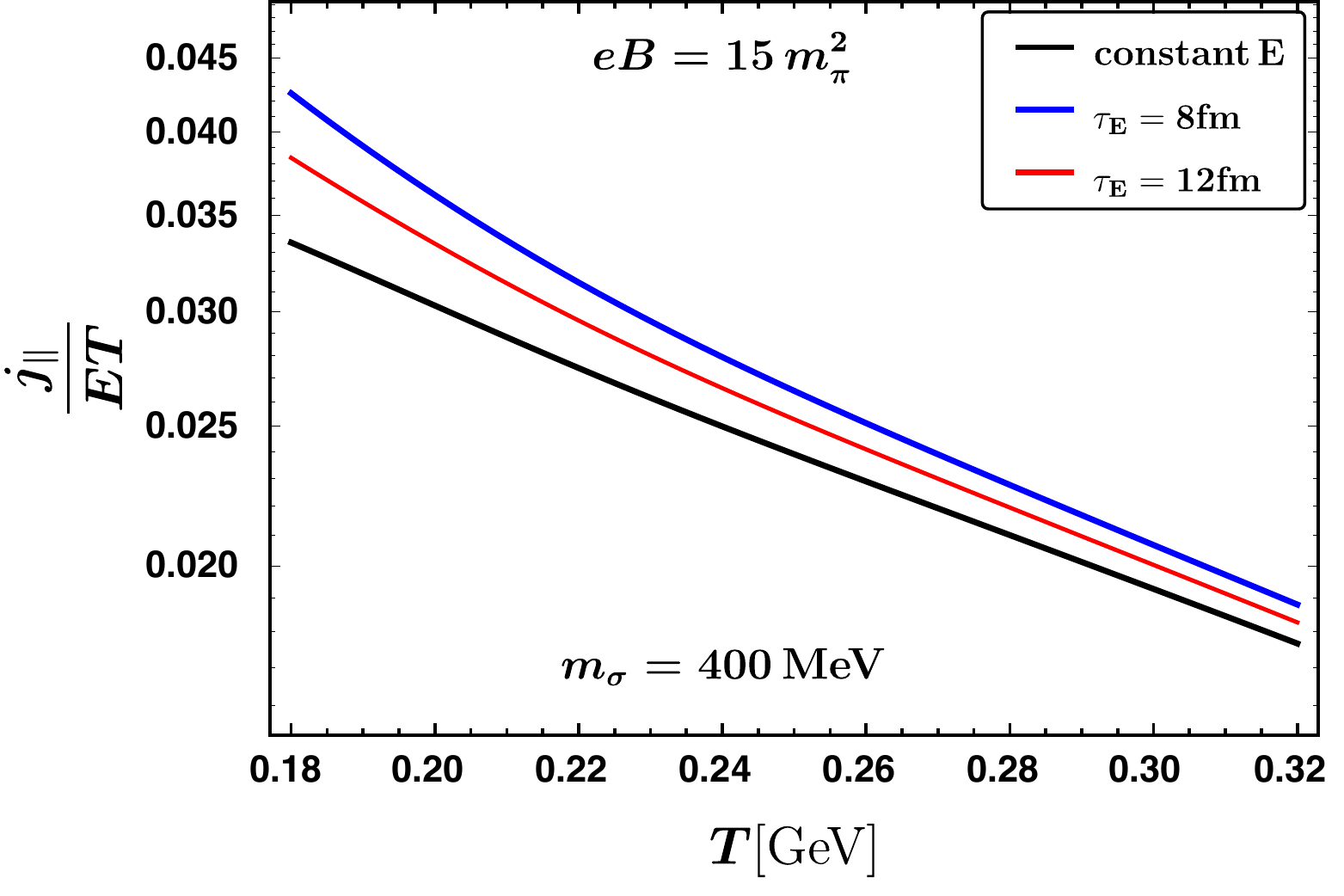}
\includegraphics[scale=0.505]{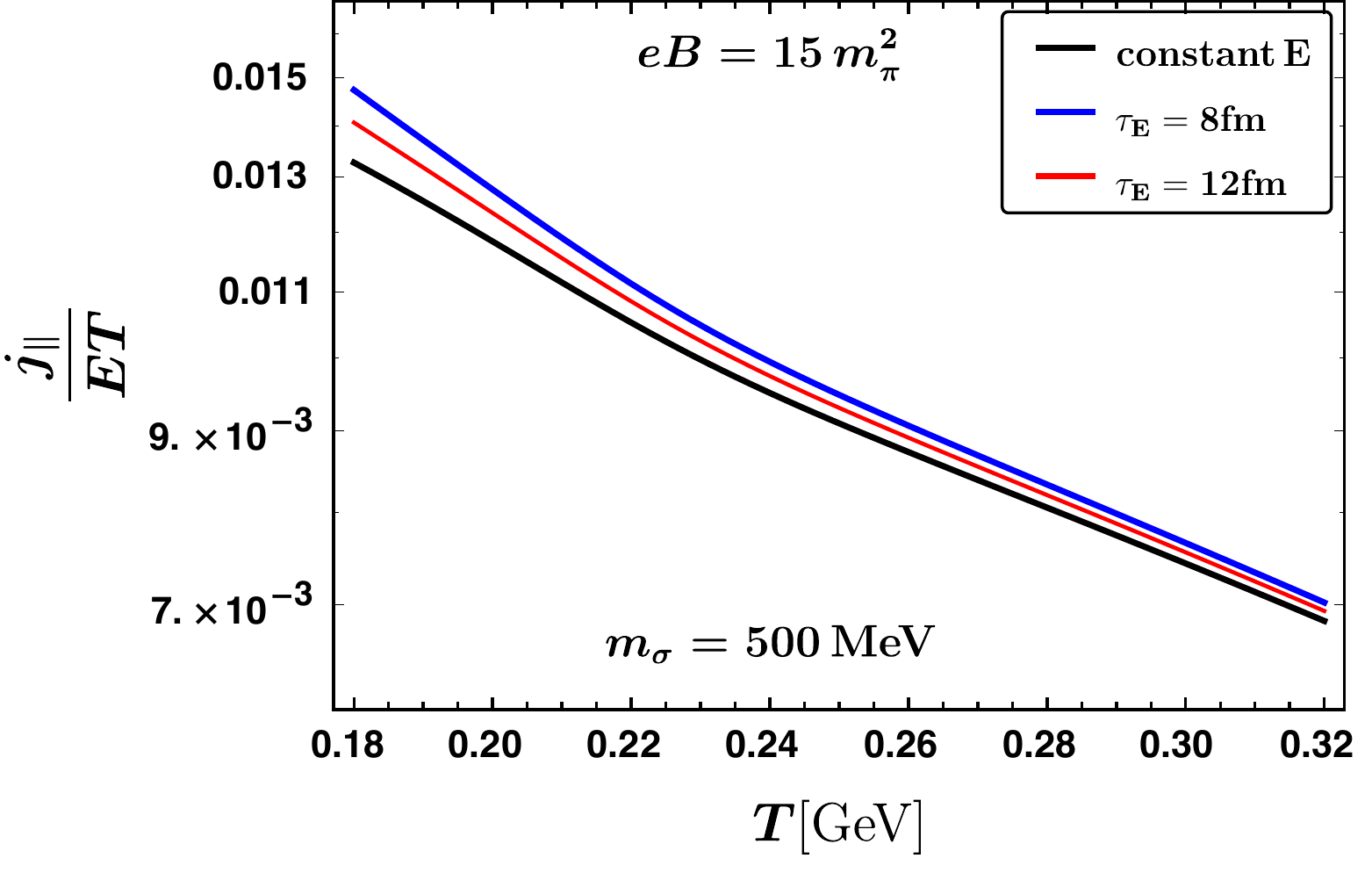}
\includegraphics[scale=0.48]{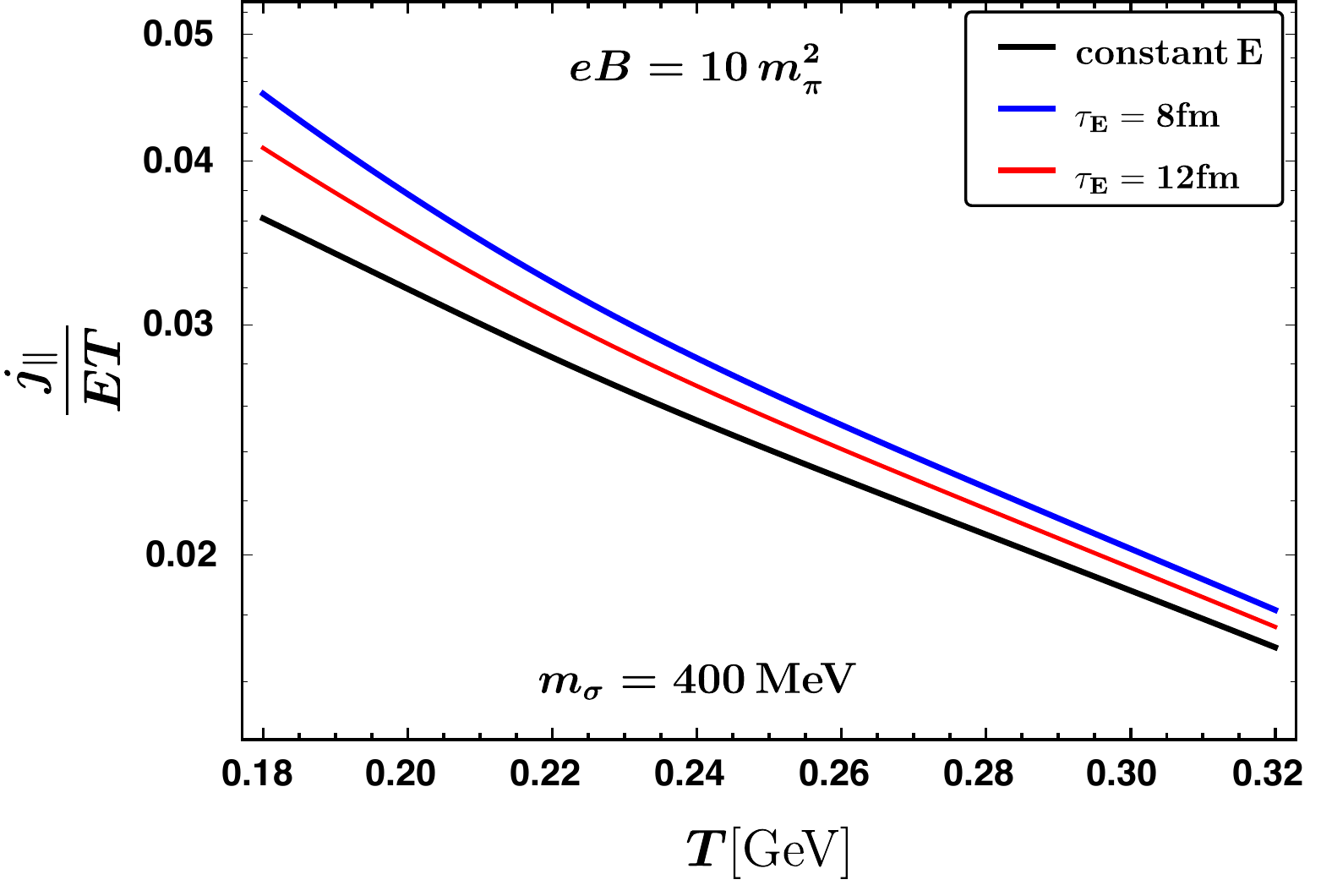}
\includegraphics[scale=0.51]{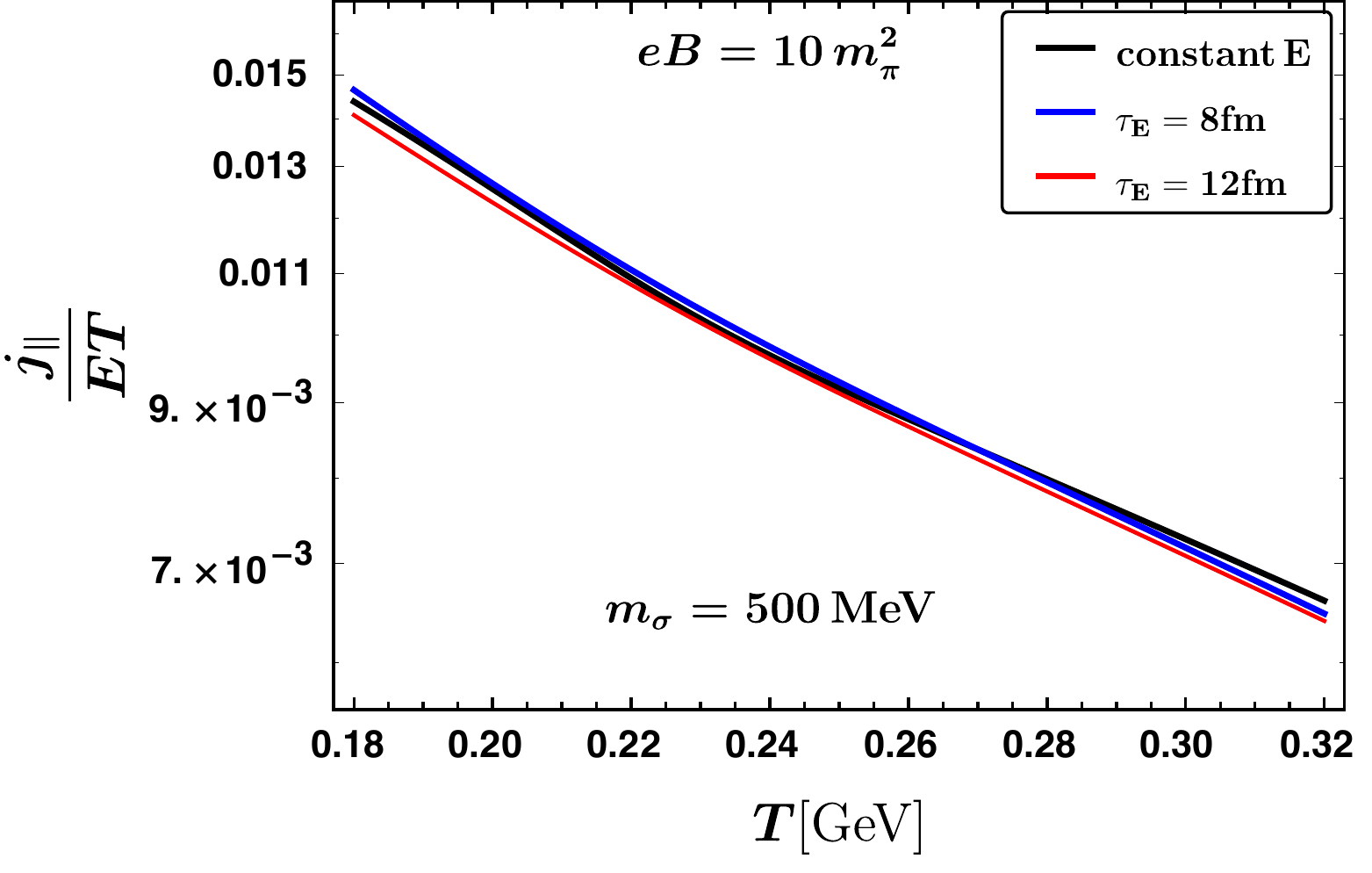}
\caption{\small Behavior of $\frac{j_{\parallel}}{ET}$  due to time-varying electric field in a strongly magnetized hadronic matter.}
\label{f6}
\end{center}
\end{figure}

The electromagnetic response of the weakly magnetized hadronic medium to a time-varying electric field is shown in Fig.~\ref{f5}.  The time dependence of the electric field significantly modifies the Ohmic current density throughout the chosen temperature range, and the effect is more pronounced for the case with lower sigma mass. Higher the value of $\tau_E$ $i.e.$, the longer the electric field stays in the medium, the smaller the impact of the time dependence of the field on the charge transport. In Fig.~\ref{f6}, the impact of the time-varying electric field on the temperature behavior of longitudinal current in a strongly magnetized hadronic matter is shown for different values of sigma mass. As the strength of the magnetic field increases (from weak field to strong-field regime), we observe a decrement in the additional contribution to the longitudinal current density in comparison with that in the weakly magnetized hadronic matter. However, the effect of the additional component is non-negligible in the low-temperature regime. Notably, at the limit $\tau_E\rightarrow \infty$, the results reduce back to that for the constant fields, both in weak and strong field regimes.

\section{Summary and outlook} \label{SU}  

We have employed a general formalism to study the electric charge transport in a magnetized hadronic matter. We have analyzed of the electric current density and associated conductivities within the LSM at finite temperature and magnetic field. Depending upon the strength of the magnetic field in the hadronic medium, the analysis has been done in weak and strong magnetic field regimes. In a weakly magnetized medium, the temperature of the medium is the dominant energy scale and magnetic field will not directly affect the thermodynamics of the medium. However, the magnetic field effects are entering through the Lorentz force term in the relativistic transport equation. On the other hand, a strong magnetic field modifies the thermodynamics via Landau level kinematics of the charged hadronic particles and interaction frequency in the medium. The sigma mass is seen to have a significant impact on the electric charge transport in the magnetized hadronic medium. We have presented the first calculations of magnetic field-dependent electrical conductivity with the LSM, both in the weak and strong field regimes. Further, the response of the hadronic medium to a time-varying external electric field is studied. We have obtained an additional component of current density due to the proper time dependence of the external field. The effect of the additional component has a strong dependence on the strength of the magnetic field and sigma mass. The evolution of the external field in the medium significantly modifies charge transport in a weakly magnetized hadronic medium, whereas the impact of field evolution is negligible in high-temperature regimes in a strongly magnetized medium.

The contribution from higher Landau levels to the electric charge transport may not be negligible with a moderate strength of the magnetic field. The formulation of interaction frequency and setting up of the LSM with higher Landau levels is an interesting aspect to explore in the near future. The magnetic field induced anisotropic transport coefficients of a chiral QCD medium is another direction to follow.

\section*{Acknowledgments}
We thank Charles Gale, Matthew Heffernan, and Han Gao for useful comments and suggestions. We acknowledge useful discussions with Najmul Haque, Vinod Chandra, and a numerical help from Gokul Krishna. This work was supported in part by the Natural Sciences and Engineering Research Council of Canada. MK acknowledges a scholarship from the Fonds de recherche du Quebec - Nature et technologies (FRQNT). 

\bibliographystyle{apsrev4-1}
\bibliography{main.bib}

\end{document}